\documentclass[twocolumn]{aastex62}
\usepackage[varg]{txfonts}
\usepackage{graphicx}
\usepackage{color}
\usepackage{eqnarray}
\begin{document}

\title{HL Tau disk in HCO$^+$ (3--2) and (1--0) with ALMA: gas density, temperature, gap, and one-arm spiral}

\author{Hsi-Wei Yen}
\affiliation{Academia Sinica Institute of Astronomy and Astrophysics, 11F of Astro-Math Bldg, 1, Sec. 4, Roosevelt Rd, Taipei 10617, Taiwan}

\author{Pin-Gao Gu} 
\affiliation{Academia Sinica Institute of Astronomy and Astrophysics, 11F of Astro-Math Bldg, 1, Sec. 4, Roosevelt Rd, Taipei 10617, Taiwan}

\author{Naomi Hirano} 
\affiliation{Academia Sinica Institute of Astronomy and Astrophysics, 11F of Astro-Math Bldg, 1, Sec. 4, Roosevelt Rd, Taipei 10617, Taiwan}

\author{Patrick M. Koch}
\affiliation{Academia Sinica Institute of Astronomy and Astrophysics, 11F of Astro-Math Bldg, 1, Sec. 4, Roosevelt Rd, Taipei 10617, Taiwan}

\author{Chin-Fei Lee}
\affiliation{Academia Sinica Institute of Astronomy and Astrophysics, 11F of Astro-Math Bldg, 1, Sec. 4, Roosevelt Rd, Taipei 10617, Taiwan}

\author{Hauyu Baobab Liu}
\affiliation{Academia Sinica Institute of Astronomy and Astrophysics, 11F of Astro-Math Bldg, 1, Sec. 4, Roosevelt Rd, Taipei 10617, Taiwan}

\author{Shigehisa Takakuwa}
\affiliation{Department of Physics and Astronomy, Graduate School of Science and Engineering, Kagoshima University, 1-21-35 Korimoto, Kagoshima, Kagoshima 890-0065, Japan}
\affiliation{Academia Sinica Institute of Astronomy and Astrophysics, 11F of Astro-Math Bldg, 1, Sec. 4, Roosevelt Rd, Taipei 10617, Taiwan}

\correspondingauthor{Hsi-Wei Yen}
\email{hwyen@asiaa.sinica.edu.tw}

\begin{abstract}
We present our observational results of the 1.1 mm continuum and the HCO$^+$ (3--2) line in HL Tau at angular resolutions of 0\farcs1 obtained with ALMA and our data analysis of the 2.9 mm and 1.1 mm continuum and the HCO$^+$ (3--2) and (1--0) lines of the HL Tau disk. The Keplerian rotation of the HL Tau disk is well resolved in the HCO$^+$ (3--2) emission, and the stellar mass is estimated to be 2.1$\pm$0.2 $M_\sun$ with a disk inclination angle of 47$\arcdeg$. The radial profiles of the HCO$^+$ column density and excitation temperature are measured with the LTE analysis of the two transitions of the HCO$^+$ emission. An HCO$^+$ gas gap at a radius of 30 au, where the column density drops by a factor of 4--8, is found in the HCO$^+$ column density profile, coincident with the dust gap traced by the continuum emission. 
No other clear HCO$^+$ gas gap is seen. This HCO$^+$ gas gap can be opened by a planet with mass of 0.5--0.8 $M_{\rm J}$, 
which is comparable to the planet mass adopted in numerical simulations to form the dust gap at the same radius in the HL Tau disk. In addition to the disk component, a one-arm spiral with a length of $\sim$3$\arcsec$ (520 au) stretching out from the inner disk is observed in the HCO$^+$ (3--2) emission. The observed velocity structures along the spiral suggest an infalling and rotating gas stream toward the inner disk. 
\end{abstract}

\keywords{protoplanetary disks - Stars: formation - ISM: kinematics and dynamics - ISM: individual objects (HL Tau)}

\section{Introduction}
With high sensitivity and angular resolutions of $\lesssim$0\farcs1,
the Atacama Large Millimeter/submillimeter Array (ALMA) has found rings and gaps in the continuum emission in a few tens of protoplanetary disks \citep[e.g.,][]{Long18, Andrews18}.
The presence of these rings and gaps could be a signpost of forming planets in protoplanetary disks, 
where the gaps are opened by gas giant planets \citep[e.g.,][]{Dong15, Dipierro15, Picogna15, Jin16, Boley17}.
Such rings and gaps in the continuum emission could also be formed by other mechanisms, such as secular gravitational instability \citep{Takahashi16, Tominaga18}, changes in dust properties and grain sizes \citep{Zhu12, Zhang15, Okuzumi16}, and the effects of magnetic field reconnection \citep{Suriano18} in protoplanetary disks.
The continuum emission traces dust, while the major mass content in protoplanetary disks is gas \citep{Williams11}. 
To understand the dynamics in protoplanetary disks and thus to distinguish these different mechanisms of formation of rings and gaps, 
observations to reveal distributions and motions of gas in those disks with continuum rings and gaps are required. 

HL Tau is a Class I--II protostar \citep{Men'shchikov99, Motte01, Furlan08} surrounded by a protoplanetary disk with a series of concentric rings and gaps in the (sub-)millimeter continuum emission \citep{ALMA15, Akiyama16}.
HL Tau is located in the northern part of the L1551 region in the Taurus molecular cloud at a distance of $\sim$140 pc \citep{Hayashi09, Galli18}, 
and it is still embedded in an infalling and rotating flattened envelope with a size of 2000--3000 au \citep{Hayashi93, Welch00, Yen17, Wu18}.
In Yen et al.~(2016a; hereafter \citet{Yen16a}), 
the analysis with azimuthal average of the ALMA archival data of the HCO$^+$ (1--0) emission in the HL Tau disk at an angular resolution of 0\farcs1 hints at two gas gaps at radii of 32 au and 69 au, coincident with the continuum gaps.
The presence of coincident gas and dust gaps suggests the actual deficit of material at these radii, and could favor the scenario where these gaps are opened by gas giant planets.
Thus, HL Tau is a promising candidate of ongoing planet formation and an excellent target to study planet formation at the early stage of star formation, when protoplanetary disks are still embedded in infalling envelopes.

Nevertheless, the gas distribution in the HL Tau disk could so far not be directly imaged at a high angular resolution of $\lesssim$0\farcs1 because of the limited signal-to-noise ratio (S/N) of the previous data, and the gas kinematics in the HL Tau disk could not be well resolved \citep{Yen16a}.
The molecular-line images of the HL Tau disk were only obtained at lower angular resolutions of $\gtrsim$0\farcs2 from the ALMA data with {\it uv}-tapering, where the signatures of rings and gaps in the gas component (if present) are smoothed out \citep{ALMA15, Pinte16, Wu18}.
If there are asymmetric structures in the disk or contaminations from the surrounding envelope, the analysis with the azimuthal average in \citet{Yen16a} could be biased.
Therefore, the presence or absence of the HCO$^+$ gas gaps in the HL Tau disk is still uncertain.

In order to directly image the gas distribution and kinematics in the HL Tau disk, 
we conducted ALMA observations at an angular resolution of 0\farcs1 in the HCO$^+$ (3--2) line.
With a typical temperature of 20 K or a few tens of K in protoplanetary disks,
the HCO$^+$ (3--2) emission is expected to be brighter than the HCO$^+$ (1--0) emission by a factor of a few.
Thus, observations in the HCO$^+$ (3--2) line can have better S/N. 
Furthermore, by adding the data of the HCO$^+$ (3--2) line, 
we are able to derive the excitation temperature and column density of the HCO$^+$ gas from the analysis of the two transitions of the HCO$^+$ emission on the assumption of the local thermal equilibrium (LTE) condition, 
leading to more robust estimates of depths of gas gaps (if present).

In this paper, we present our observational results of the HCO$^+$ (3--2) emission in the HL Tau disk obtained with ALMA and our analysis on the data of the 2.9 mm and 1.1 mm continuum and the HCO$^+$ (3--2) and (1--0) lines at angular resolutions of 0\farcs1.
The details of our ALMA observations are described in Section \ref{ob}, 
and the observational results are presented in Section \ref{result}.
In Section \ref{analysis}, we present our study of the gas kinematics traced by the HCO$^+$ (3--2) emission in the HL Tau disk, 
and our measured radial profiles of the HCO$^+$ column density and excitation temperature. 
Our analysis suggests the presence of one HCO$^+$ gas gap in the HL Tau disk and an infalling one-arm spiral connected with the disk. 
We discuss the origins of the HCO$^+$ gas gap and the infalling spiral found in our observations in Section \ref{discuss}.

\begin{table*}
\caption{Summary of images}\label{mapinfo}
\centering
\begin{tabular}{ccccc}
\hline\hline
Line/Continuum & Rest frequency & Beam size (PA) & Velocity resolution & Noise  \\
 & (GHz) & & (km s$^{-1}$) & (per beam) \\
\hline\hline
2.9 mm continuum & 102.9 & 0\farcs09 $\times$ 0\farcs06 (9\arcdeg) & \nodata & 15 $\mu$Jy \\
1.1 mm continuum & 260.7 & 0\farcs09 $\times$ 0\farcs09 (19\arcdeg) & \nodata & 30 $\mu$Jy \\
HCO$^+$ (1--0) & 89.188526 & 0\farcs1 $\times$ 0\farcs06 (9\arcdeg) & 0.21 & 1.5 mJy\\
HCO$^+$ (3--2) & 267.557633 & 0\farcs1 $\times$ 0\farcs09 (19\arcdeg) & 0.1 & 2.5 mJy\\
\hline\hline
\end{tabular}
\tablecomments{For molecular-line images, the noise is in units of per beam per channel.}
\end{table*}

\begin{figure*}
\centering
\includegraphics[width=\textwidth]{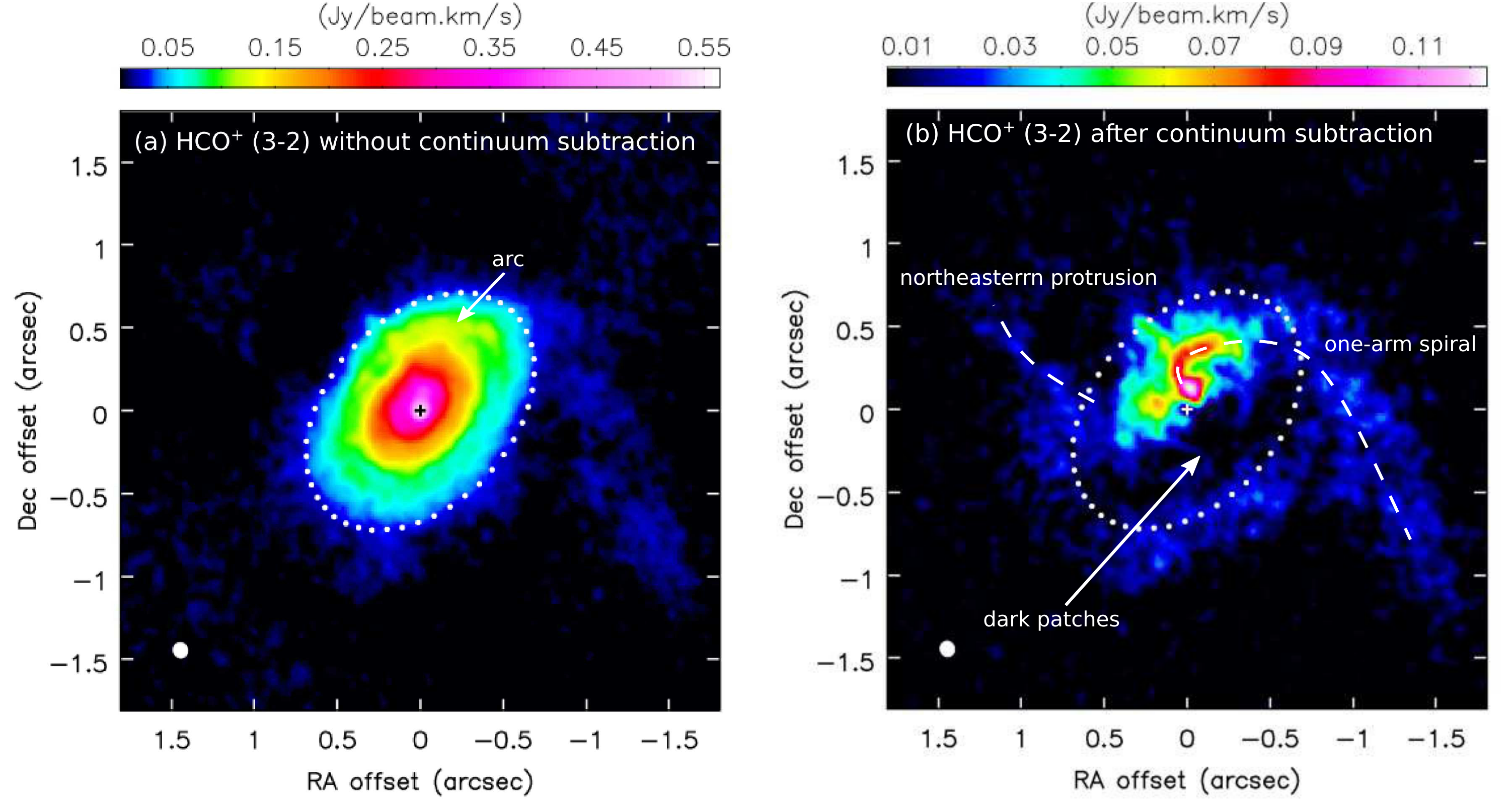}
\caption{Total integrated intensity maps in units of Jy beam$^{-1}$ km s$^{-1}$ of the HCO$^+$ (3--2) emission without (a) and with (b) the continuum subtraction in HL Tau obtained with our ALMA observations. Crosses denote the position of HL Tau. White dots delineate the disk with a radius of 0\farcs82 around HL Tau observed in the 1.1 mm continuum emission. Filled white ellipses at the bottom left corner present the beam size. 
The features discussed in the text, arc, dark patches, one-arm spiral, and northeastern protrusion, are labeled in the figures.}
\label{hcop_mom}
\end{figure*}

\section{Observations}\label{ob}
\subsection{1.1 mm observation}
The ALMA observations at 1.1 mm of HL Tau were executed on 2017 August 22 and 28 and September 1. 
The array configuration was C40-7. 
The maximum projected baseline length of the observations is $\sim$3.3 km.
On August 22, J0237+2848, J0423$-$0120, and J0431+1731 were observed as bandpass, flux, and phase calibrators, respectively.
On August 28 and September 1, J0510+1800 was observed as bandpass and flux calibrators, and J0431+1731 as a phase calibrator. 
The spectral setup consists of seven spectral windows.
Two spectral windows with a bandwidth of 2 GHz each were assigned to the continuum emission.
One spectral window with a bandwidth of 59 MHz and a native spectral resolution of 61 kHz after the Hanning smoothing was assigned to the HCO$^+$ (3--2) line, resulting in a velocity resolution of 0.068 km s$^{-1}$.
HCN (3--2), HC$^{18}$O$^+$ (3--2), and SO$_2$ (4$_{3,1}$--4$_{2,2}$; 6$_{3,3}$--6$_{2,4}$) were also observed simultaneously at high spectral resolutions of 61--122 kHz in the other spectral windows. 
In this paper, we present the results of the 1.1 mm continuum and the HCO$^+$ (3--2) line.

The data were calibrated with the pipeline of the software Common Astronomy Software Applications \citep[CASA;][]{McMullin07} of version 4.7.0. 
The calibration script obtained from the archive includes additional manual flagging to remove the data of several antennas showing phase jumps. 
We further performed self-calibration on the phase with the continuum data, 
and applied the solution to the molecular-line data. 
We generated images from the calibrated visibilities with the Briggs weighting with a robust parameter of 0.5, and cleaned the images, using the CASA task {\it tclean}.
The HCO$^+$ (3--2) images were generated with a channel width of 0.1 km s$^{-1}$, 
and two HCO$^+$ images cubes with and without the continuum subtraction were generated.
The continuum subtraction was performed using the CASA task {\it uvcontsub}.
The sizes of the synthesized beams of the 1.1 mm continuum and HCO$^+$ (3--2) images are 0\farcs09 $\times$ 0\farcs09 and 0\farcs1 $\times$ 0\farcs09, respectively.
The position angle (PA) of the synthesized beams is 19\degr.
The achieved noise level is 30 $\mu$Jy beam$^{-1}$ in the 1.1 mm continuum image,
and 2.5 mJy beam$^{-1}$  per channel in the HCO$^+$ (3--2) images. 

\subsection{2.9 mm observation}
The ALMA 2.9 mm data of HL Tau were retrieved from the public archive of the ALMA science verification data (project code: 2011.0.00015.SV).
The details of the observations and the data reduction are described in \citet{ALMA15}.
The 2.9 mm continuum image analyzed in this work was directly obtained from the archive. 
The imaging process of this continuum image is described in the CASA guide\footnote{\url{https://casaguides.nrao.edu/index.php?title=ALMA2014_LBC_SVDATA}}.
The 2.9 continuum image has a synthesized beam with a size of 0\farcs09 $\times$ 0\farcs06 and a PA of 1\degr, 
and its noise level is 15 $\mu$Jy beam$^{-1}$.
We retrieved the calibrated visibility data of the HCO$^+$ (1--0) line from the archive, 
and subtracted the continuum from the HCO$^+$ (1--0) data using the CASA task {\it uvcontsub}. 
Then we generated HCO$^+$ (1--0) images with a channel width of 0.21 km s$^{-1}$ and cleaned the images using the CASA task {\it tclean}. 
The Briggs weighting with a robust parameter of 0 and without {\it uv} tapering was adopted, resulting in a synthesized beam with a size of 0\farcs1 $\times$ 0\farcs06 and a PA of 9\degr. 
The achieved noise level is 1.5 mJy beam$^{-1}$ per channel.

\begin{figure*}
\centering
\includegraphics[width=\textwidth]{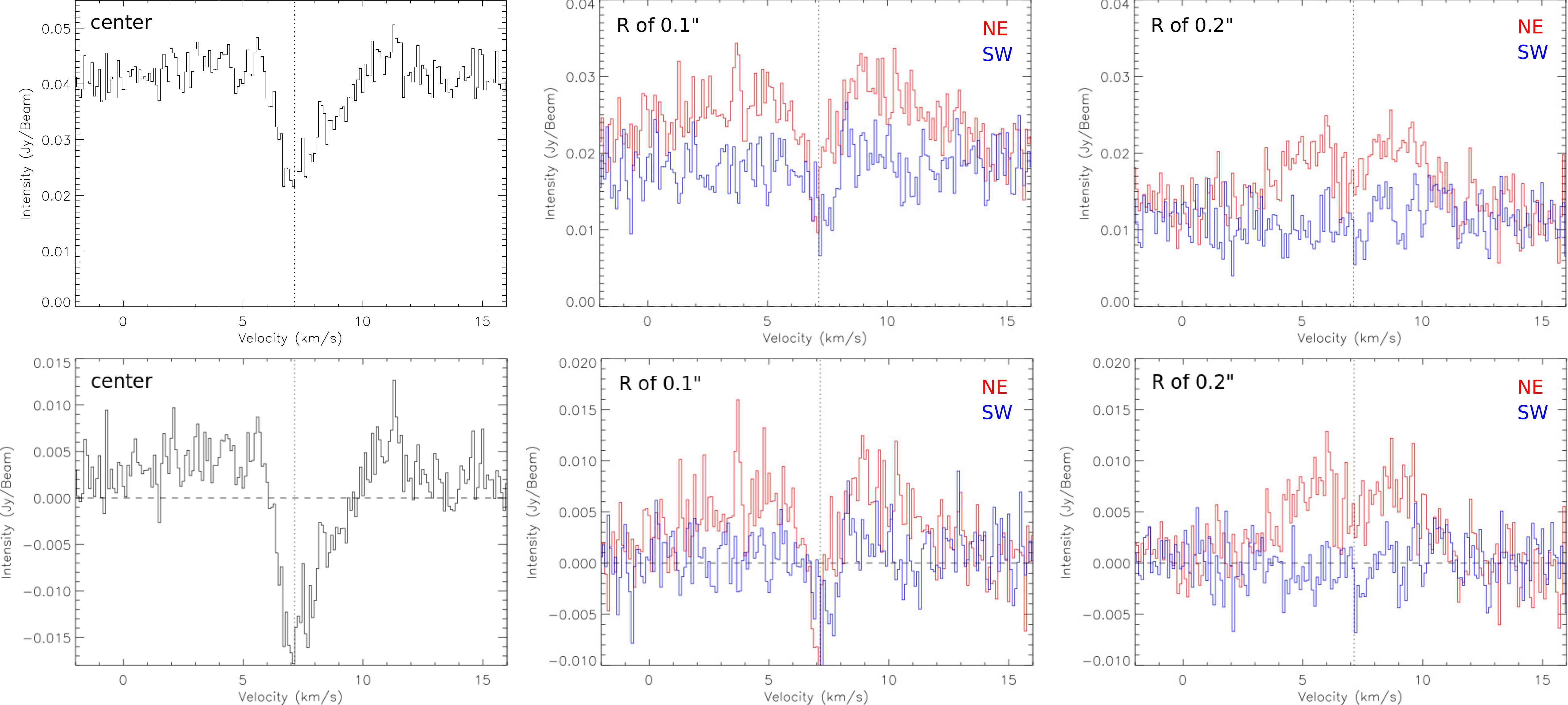}
\caption{HCO$^+$ (3--2) spectra without (upper) and with (lower) the continuum subtraction at different positions, center (left column) and a projected radius of 0\farcs1 (middle column) and 0\farcs2 (right column) along the minor axis. Blue and red histograms present the spectra extracted from the northeastern and southwestern sides of the disk, respectively. Dotted vertical lines denote the systemic velocity of HL Tau (Section~\ref{MsVsys}). Horizontal dashed lines denote zero intensity.}
\label{hcop_spec}
\end{figure*}

\section{Results}\label{result}
The rest frequencies, angular and velocity resolutions, and noise levels of all the images analyzed in this work are summarized in Table \ref{mapinfo}.
The results of the HCO$^+$ (1--0) and continuum emission at millimeter wavelengths obtained with ALMA have been presented and discussed in details in the literature \citep[e.g.,][]{ALMA15, Akiyama16, Pinte16, Yen16a, Wu18}.
Thus, in this section, we only present the results of the HCO$^+$ (3--2) line.
The 1.1 mm continuum image obtained with our observations is shown in Appendix \ref{1.1mm_cont}.
The analysis combining all four images is presented in Section~\ref{TrNr}.

\subsection{HCO$^+$ intensity distribution}\label{hcop32}

Figure~\ref{hcop_mom} presents the total integrated intensity (moment 0) maps of the HCO$^+$ (3--2) emission without and with the continuum subtraction.
In the moment 0 map without the continuum subtraction, 
the size of the overall emission distribution is comparable to the disk size observed in the 1.1 mm continuum 
because the emission is dominated by the continuum flux.
The emission is brighter in the northeastern side than in the southwestern side in the disk.
In addition, an arc feature is seen in the north, and it extends to the west and then to the southwest. 
After the continuum subtraction to show the line emission in excess of the continuum flux, 
the asymmetric intensity distribution and the arc feature are clearly seen. 
In the moment 0 map with the continuum subtraction,
there are dark patches in the southwestern part of the disk,
and a bright peak located to the north of HL Tau. 
A spiral structure stretches out from this peak position toward the north, then bends toward the west, 
and further extends to the southwest 2$\arcsec$ away from the center. 
The position of the bright peak is measured to be ($-0\farcs02$, 0\farcs12) with respect to the position of HL Tau. 
If this peak is located in the disk mid-plane, then its de-projected radius is 20 au. 
There is an additional protrusion seen in the northeast in the moment 0 map with the continuum subtraction.

Figure~\ref{hcop_spec} compares the spectra of the HCO$^+$ (3--2) line without and with the continuum subtraction at different positions in the disk.
The continuum flux is relatively high compared to that of the line emission in excess of the continuum. 
Before the continuum subtraction, the spectrum at the center shows a clear absorption feature centered at a velocity of $\sim$7 km s$^{-1}$.
Similar absorption features are also seen in the spectra extracted at a radius of 0\farcs1 along the disk minor axis with a PA of 48$\arcdeg$ in both northeastern and southwestern sides of the disk. 
In addition, the line emission is brighter in the northwest than in the southeast by a factor of 1.5--2.
After the continuum subtraction, these absorption features show negative intensities. 
Similar absorption has also been observed in the HCO$^+$ (1--0), CN (1--0), and HCN (1--0) lines with ALMA \citep{ALMA15}.
Furthermore, after the continuum subtraction, there is almost no emission in the spectra extracted at radii of 0\farcs1--0\farcs2 in the southwest along the disk minor axis.
Thus, the dark patches in the moment 0 map of the HCO$^+$ emission with the continuum subtraction are due to the absorption and fainter emission in the southwestern part of the disk (Fig.~\ref{hcop_mom}b).

\begin{figure*}
\centering
\includegraphics[width=\textwidth]{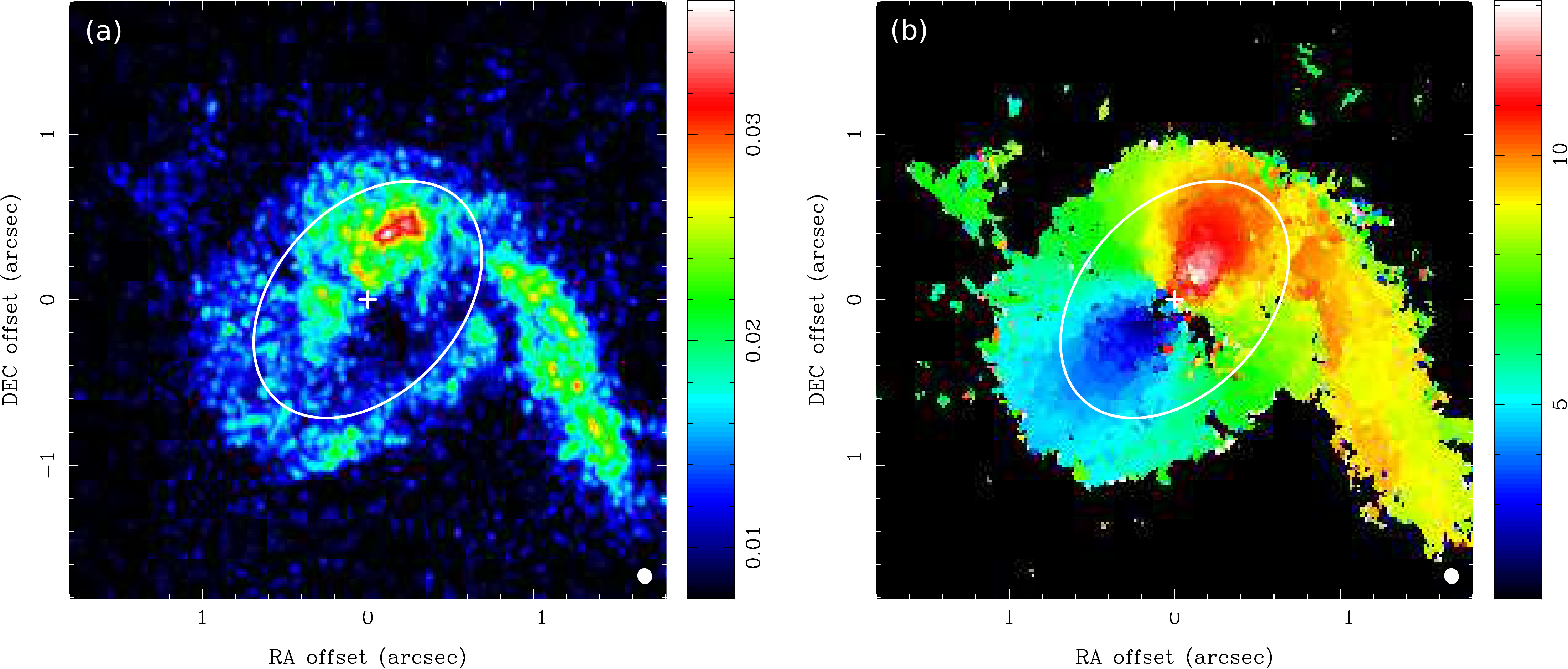}
\caption{Maps of (a) the peak intensities (moment 8) in units of Jy beam$^{-1}$ and (b) the velocities of the intensity peaks (moment 9) in units of km s$^{-1}$ in the LSR frame of the HCO$^+$ (3--2) emission after the continuum subtraction. Crosses denote the position of HL Tau. White open ellipses delineate the size of the disk around HL Tau observed in the 1.1 mm continuum. White filled ellipses at the bottom right corners present the beam size.}
\label{hcop_mom1}
\end{figure*}

To show the complete spatial distribution of the HCO$^+$ (3--2) emission more clearly, 
Fig.~\ref{hcop_mom1}a presents the peak intensity (moment 8) map generated from the HCO$^+$ image cube with the continuum subtraction. 
The HCO$^+$ (3--2) emission is detected above 3$\sigma$ in the most part of the disk,
and the apparent size of the intensity distribution is larger than that of the continuum disk.
Overall the peak intensities in the northeast are higher than in the southwest in the disk.
There is a hole in the southwest close to the center in the moment 8 map, 
where there is no clear detection above 3$\sigma$.
The spiral structure extending from the inner disk to the southwest is clearly seen in the moment 8 map, 
and the peak intensities along the spiral structure are higher than those in the most parts of the disk.

\subsection{Velocity structure}

Figure \ref{hcop_mom1}b presents the map of the velocities of the intensities peaks (moment 9) of the HCO$^+$ (3--2) emission generated from the image cube with the continuum subtraction.
In the moment 9 map, the HCO$^+$ line exhibits a velocity gradient along the disk major axis of a PA of 138$\arcdeg$, and the southeast and the northwest are blue- and redshifted, respectively.
This velocity gradient has also been observed in the HCO$^+$ (1--0) emission at angular resolutions of 0\farcs25--1$\arcsec$ \citep{ALMA15, Pinte16, Wu18}. 
There is no clear velocity gradient along the disk minor axis, where is greenish in the both northeast and southwest in the disk in the moment 9 map. 
The spiral structure is redshifted, and there is likely a velocity gradient along the spiral structure, where the emission is less redshifted at outer radii.

\begin{figure*}
\centering
\includegraphics[width=17cm]{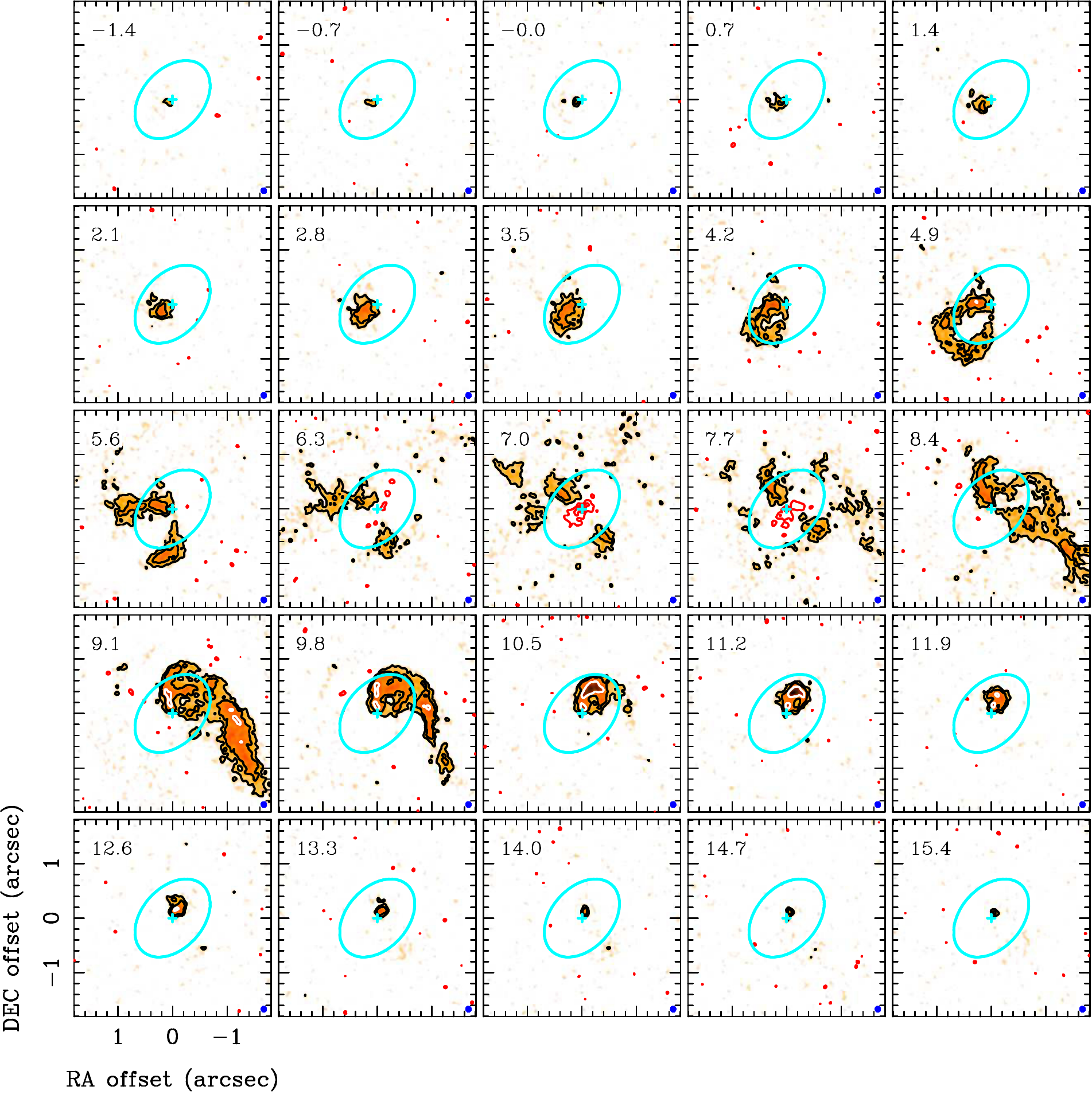}
\caption{Velocity channel maps of the HCO$^+$ (3--2) emission in HL Tau obtained with our ALMA observations. The central velocity of each channel in units of km s$^{-1}$ is labeled at the upper left corner in each panel. Cyan crosses denote the position of HL Tau. Cyan open ellipses present the size of the continuum disk with a radius of 0\farcs82 around HL Tau observed at 1.1 mm. Contour levels are 5$\sigma$, 10$\sigma$, 20$\sigma$ and 40$\sigma$, where 1$\sigma$ is 0.9 mJy beam$^{-1}$. Negative contours are shown in red and are $-4\sigma$ and $-8\sigma$. A blue ellipse at the lower right corner in each panel presents the size of the synthesized beam.}
\label{hcop_chan}
\end{figure*}

Figure \ref{hcop_chan} presents the velocity channel maps of the HCO$^+$ (3--2) emission after the continuum subtraction. 
To reduce the number of panels for better presentation, 
the velocity channel maps are binned up every seven channels to have a velocity resolution of 0.7 km s$^{-1}$. 
The original velocity channel maps with a velocity resolution of 0.1 km s$^{-1}$ were used for analysis.
The velocity pattern of the disk rotation is clearly seen in the HCO$^+$ (3--2) emission. 
The emission is elongated along the minor axis of the disk at $V_{\rm LSR}$ of 7.0--7.7 km s$^{-1}$.
Thus, the systemic velocity ($V_{\rm sys}$) of the HL Tau disk is expected to be between 7.0 km s$^{-1}$ and 7.7 km s$^{-1}$. 
At these velocities close to $V_{\rm sys}$, there is significant negative intensity at a level of more than 3$\sigma$ (red contours), 
and the peak negative intensity is 17$\sigma$, as seen in the spectra extracted around the center (Fig.~\ref{hcop_spec}).
The emission is compact and located close to the center at a higher relative velocity with respect to $V_{\rm sys}$, 
and becomes more extended at a lower relative velocity.  
At $V_{\rm LSR}$ of 4.2--4.9 km s$^{-1}$, the emission exhibits a ring-like structure.
These observed features are consistent with the expectation from Keplerian rotation \citep[e.g., see Fig.~8 in][]{Rosenfeld13}.
The redshifted emission at $V_{\rm LSR}$ of 7.7--9.8 km s$^{-1}$ additionally shows a spiral structure with a length of $\sim$3$\arcsec$ (520 au) connecting to the northwestern side of the disk.
The spiral structure is more extended at a lower relative velocity of $V_{\rm LSR}$ of 9.1 km s$^{-1}$ than at a higher velocity of $V_{\rm LSR}$ of 9.8 km s$^{-1}$, suggesting a velocity gradient along the spiral structure, as seen in the moment 9 map.

\begin{figure*}
\centering
\includegraphics[width=\textwidth]{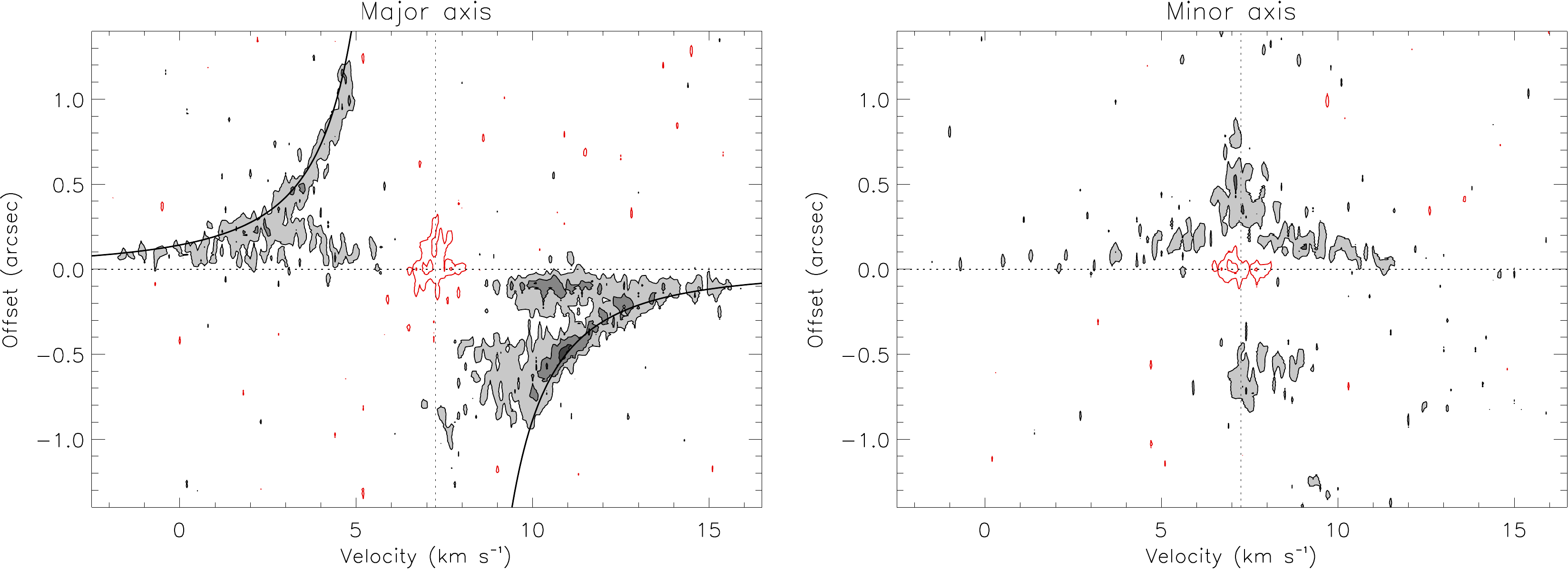}
\caption{Position--velocity diagrams of the HCO$^+$ (3--2) emission along the major (left) and minor (right) axes of the disk in HL Tau. Dotted horizontal and vertical lines denote the position of HL Tau and the systemic velocity of $V_{\rm LSR}$ of 7.14 km s$^{-1}$, respectively. The solid curves present the expected velocity profile of Keplerian rotation around a central stellar mass of 2.14 $M_\sun$ with an inclination angle of the disk mid-plane of 47$\arcdeg$. Contour levels are from 3$\sigma$ in steps of 3$\sigma$, where 1$\sigma$ is 2.5 mJy beam$^{-1}$. Negative contours are plotted in red color.}
\label{hcop_pv}
\end{figure*}

Figure \ref{hcop_pv} presents the position--velocity (PV) diagrams of the HCO$^+$ (3--2) emission along the major and minor axes of the disk in HL Tau.
The PV diagram along the major axis clearly shows a signature of spin-up rotation, where the velocity of the emission is higher at a smaller radius. 
In addition, the velocity profile is consistent with that of Keplerian rotation, $V \propto R^{-0.5}$ (solid curves). 
Especially at the blueshifted velocity, the emission at outer radii with an offset $>0\farcs3$ follows the expected profile of the Keplerian rotation. 
On the other hand, at the redshifted velocity, there is additional emission detected at a velocity slower than the expected Keplerian velocity at outer radii with offsets from $-0\farcs4$ to $-0\farcs8$, {which could be the contamination from the ambient gas}.
Along the minor axis, 
there is no clear velocity gradient in the PV diagram.
These PV diagrams suggest that the gas motion traced by the HCO$^+$ (3--2) emission is dominated by the disk rotation \citep[e.g.,][]{Ohashi97}.

\subsection{HCO$^+$ absorption}

\begin{figure*}
\centering
\includegraphics[width=\textwidth]{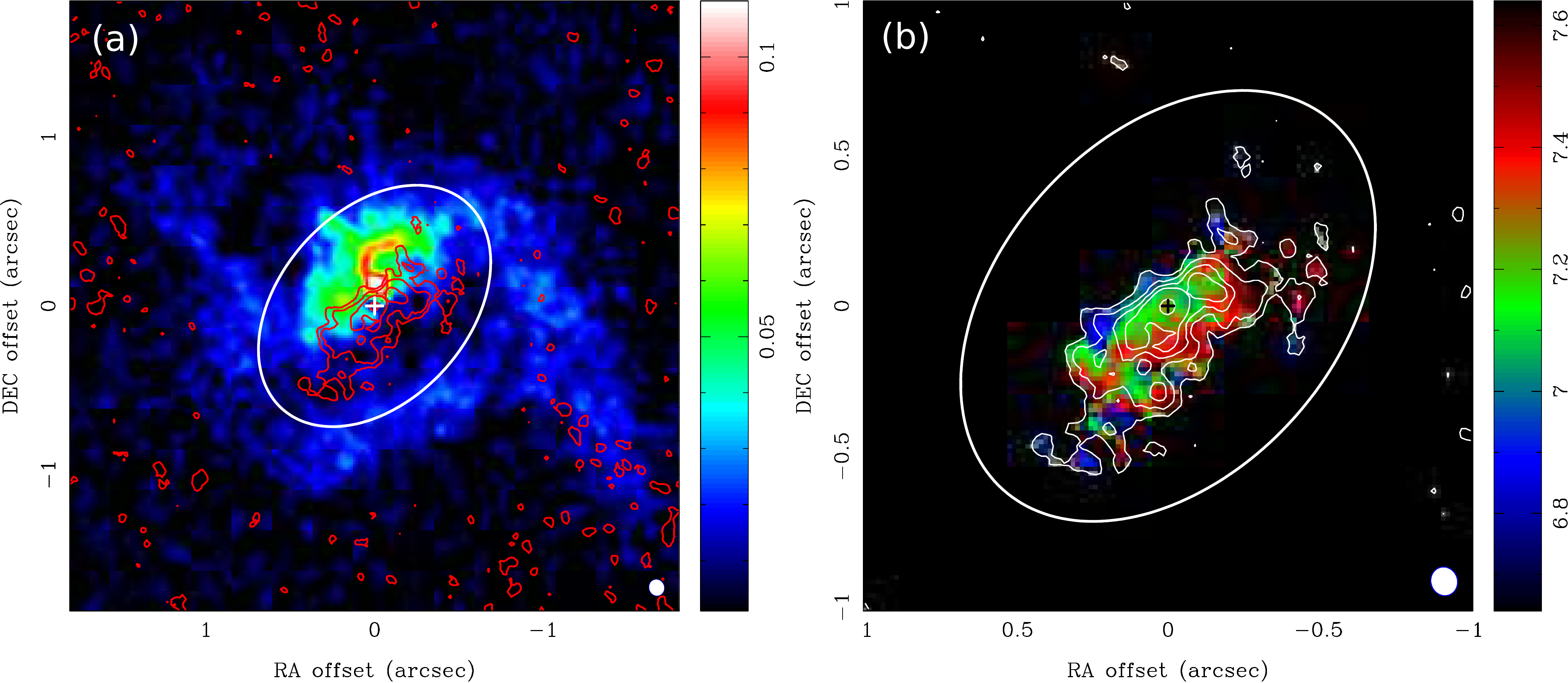}
\caption{(a) Comparison of the moment 0 maps of the emission (color scale) and absorption (red contours) of the HCO$^+$ (3--2) line observed with ALMA. The moment 0 map of the emission is the same as Fig.~\ref{hcop_mom}b. The moment 0 map of the absorption is generated by integrating the velocity range from $V_{\rm LSR}$ of 6.6 km s$^{-1}$ to 7.9 km s$^{-1}$, where the absorption is detected at a level deeper than $-3\sigma$ in the velocity channel maps. Contour levels are $-2\sigma$, $-4\sigma$, $-8\sigma$, and $-16\sigma$, where 1$\sigma$ is 1 mJy beam$^{-1}$ km s$^{-1}$. (b) Intensity weighted mean velocity (moment 1) map of the absorption (color scale) in units of km s$^{-1}$ in the LSR frame. Contours are the same as in (a). Crosses denote the position of HL Tau, and open ellipses delineate the disk with a radius of 0\farcs82 observed in the 1.1 mm continuum. Filled white ellipses at the bottom right corners present the beam size.}
\label{hcop_absorption}
\end{figure*}

Figure \ref{hcop_absorption}a compares the moment 0 maps of the HCO$^+$ (3--2) emission and absorption.
The HCO$^+$ (3--2) absorption is elongated along the northwest--southeast direction, similar to the dark patches in the moment 0 map, 
and it is deeper closer to the center. 
As shown in Fig.~\ref{hcop_absorption}b, the absorption is mostly redshifted, 
and there is a hint of a velocity gradient along the disk minor axis in the absorption. 
On the other hand, 
the direction of the velocity gradient due to the disk rotation is along the disk major axis.   
The velocity structure of the absorption does not follow the disk rotation.
Thus, the absorption is unlikely caused by cold gas in the disk, 
and it is more likely caused by the surrounding infalling envelope on the near side, which is at the redshifted velocity \citep{Yen17, Yen18, Wu18}.
The IRAM 30m and ALMA observations in the HCO$^+$ (1--0) line show that there is indeed ambient gas around the disk over a wide velocity range from $V_{\rm LSR}$ of 2.5 km s$^{-1}$ to 11.5 km s$^{-1}$ \citep{Wu18}.

\section{Analysis}\label{analysis}

\begin{figure}
\centering
\includegraphics[width=8.5cm]{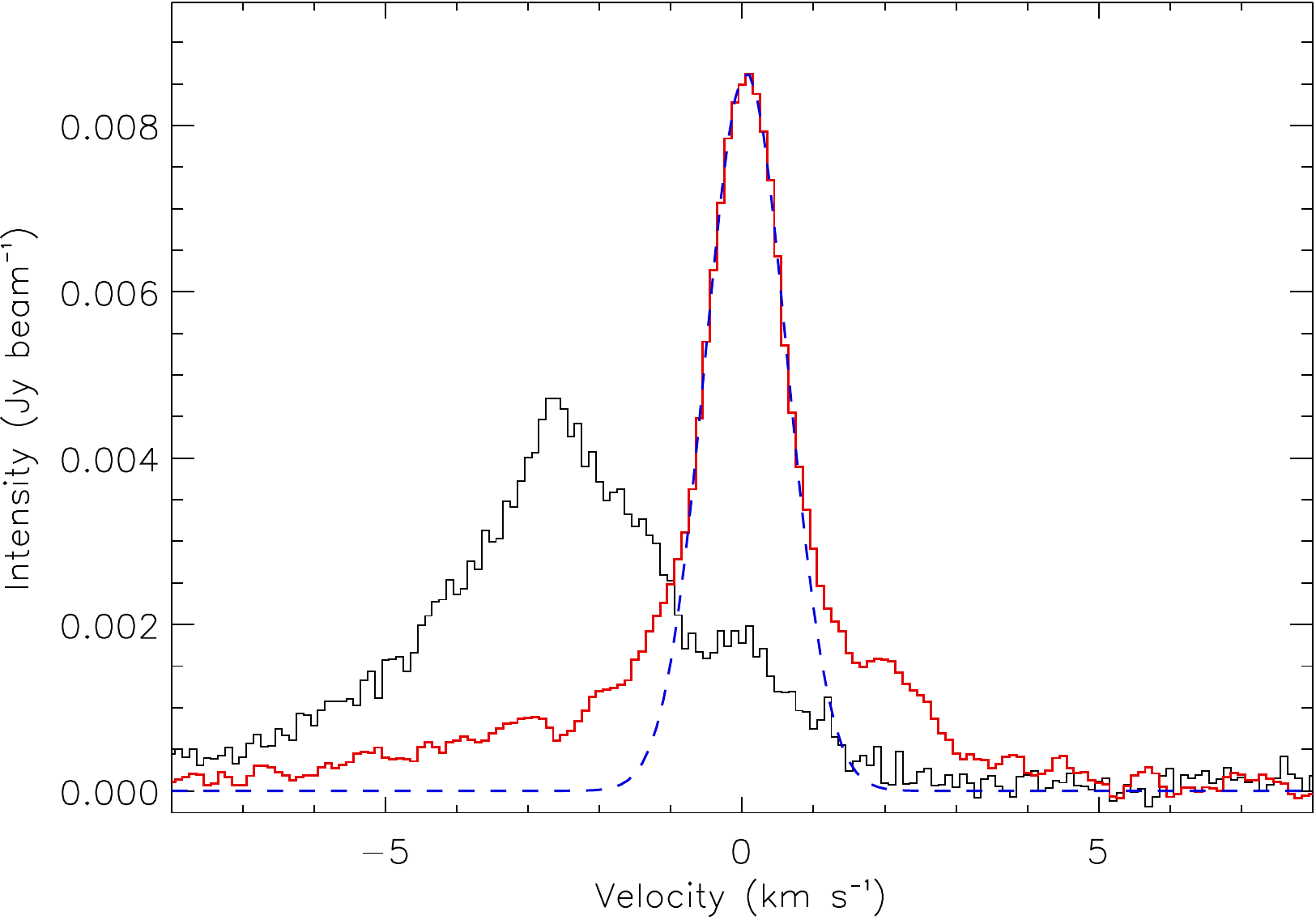}
\caption{Averaged spectrum of the blueshifted side of the disk (black histogram) and the best velocity-aligned stacked spectrum of the same region from our method (red histogram). The parameters for the alignment are $M_\star$ of 2.14 $M_\sun$ and $V_{\rm sys}$ of 7.14 km s$^{-1}$. The velocity axis presents the relative velocity with respect to the measured $V_{\rm sys}$ of 7.14 km s$^{-1}$. A blue curve shows the fitted Gaussian profile to the central part of the velocity-aligned stacked spectrum, where the intensity is above half of the peak intensity.}
\label{aligned_spec}
\end{figure}

\subsection{Stellar mass and systemic velocity}\label{MsVsys}
Because of the asymmetric intensity distribution of the HCO$^+$ (3--2) emission in the disk around HL Tau (Fig.~\ref{hcop_mom}), 
it is not straightforward to construct a kinematical model to fit the observed velocity channel maps.
Alternatively, we adopt the velocity-aligned stacking method introduced in \citet{Yen16b, Yen18} to measure $M_\star$ and $V_{\rm sys}$.
This method aligns spectra at different positions in a disk with various velocity patterns of Keplerian rotation. 
The S/N of the velocity-aligned stacked spectra are expected to reach the maximum when the adopted velocity pattern for the alignment matches the actual disk rotation \citep{Yen16b}.
Thus, $M_\star$ and $V_{\rm sys}$ can be estimated by searching for a velocity pattern of Keplerian rotation that maximizes the S/N of the velocity-aligned stacked spectra\footnote{This method conserves the flux and narrows the line width. Thus, in our analysis, we maximized the S/N of the integrated intensity of the velocity-aligned stacked spectrum within the 1$\sigma$ Gaussian line width with respect to its centroid velocity to measure $M_\star$ and $V_{\rm sys}$. The detailed process is described in \citet{Yen18}.} \citep{Yen18}. 

Because of the strong absorption in the HCO$^+$ (3--2) line close to the center, 
which could affect our estimation, 
we follow the procedure in \citet{Yen16a} to mask the regions showing negative intensity in the velocity channel maps. 
Because of the limited S/N to identify low-level absorption in the velocity channel maps, we first generated a series of azimuthally averaged spectra in different radial bins. 
The width of the radial bins is half of the beam size. 
For each radial bin, we fitted a Gaussian profile to the absorption line profile (if present) in the azimuthally averaged spectrum and measured the line width of the absorption. 
We masked the channels within twice the 1$\sigma$ Gaussian line width with respect to the centroid velocity of the absorption in each radial bin.
In addition, the spiral structure is observed in the northwestern part of the disk. 
Its motion may not follow the Keplerian rotation of the disk.
To have minimal contamination from other motions in addition to the Keplerian rotation, 
we only selected the data of the southeastern part of the disk with PA of 48\arcdeg--228\arcdeg, i.e., the blueshifted region, for our analysis.

In our analysis, we fixed the position angle of the disk major axis and the disk inclination angle at 138$\arcdeg$ and 47$\arcdeg$, respectively, identical to those measured in the millimeter continuum image at a high angular resolution of 0\farcs03 obtained with ALMA \citep{ALMA15}.
The free parameters in the analysis are $M_\star$ and $V_{\rm sys}$. 
The velocity patterns of Keplerian rotation were computed with the assumption of a geometrically thin disk.
Figure \ref{aligned_spec} presents the spectrum averaged over the selected region directly extracted from the data cube without alignment in comparison with the velocity-aligned stacked spectrum having the maximum S/N of the same region from our method.
$M_\star$ and $V_{\rm sys}$ are estimated to be 2.14$\pm$0.11 $M_\sun$ and $V_{\rm LSR}$ of 7.14$\pm$0.03 km s$^{-1}$, respectively.
These estimated uncertainties in $M_\star$ and $V_{\rm sys}$ only include the uncertainty due to the noise in the data. 
The uncertainties are defined such that the S/N and line profiles of the resultant velocity-aligned stacked spectra are all consistent within the uncertainty \citep{Yen18}. 
The estimated $M_\star$ is higher than the previous estimate of 1.8 $M_\sun$ from the PV diagram along the disk major axis of the $^{13}$CO and C$^{18}$O emission observed with ALMA at angular resolutions of 0\farcs8 \citep{Yen17}. 
This difference could be due to the lower angular resolution of the ALMA $^{13}$CO and C$^{18}$O observations.  
As discussed in \citet{Aso15}, $M_\star$ could be underestimated from a PV diagram of Keplerian rotation observed at a lower angular resolution.
In addition, as shown in Section \ref{TrNr}, the HCO$^+$ (3--2) emission is optically thick. 
Optically thick lines cannot trace the mid-plane of the disk \citep[e.g.][]{Rosenfeld13}. 
For example, with a disk aspect ratio of 0.1 (see Section \ref{discuss1}), 
the difference between the Keplerian velocities in the mid-plane and the surface is $\sim$2\%.
Considering there is likely an additional uncertainty of a few percent in the estimated $M_\star$ due to the unknown height of the emitting layer of the HCO$^+$ (3--2) emission, 
we estimate $M_\star$ to be 2.1$\pm$0.2 $M_\odot$.

\begin{figure*}
\centering
\includegraphics[width=\textwidth]{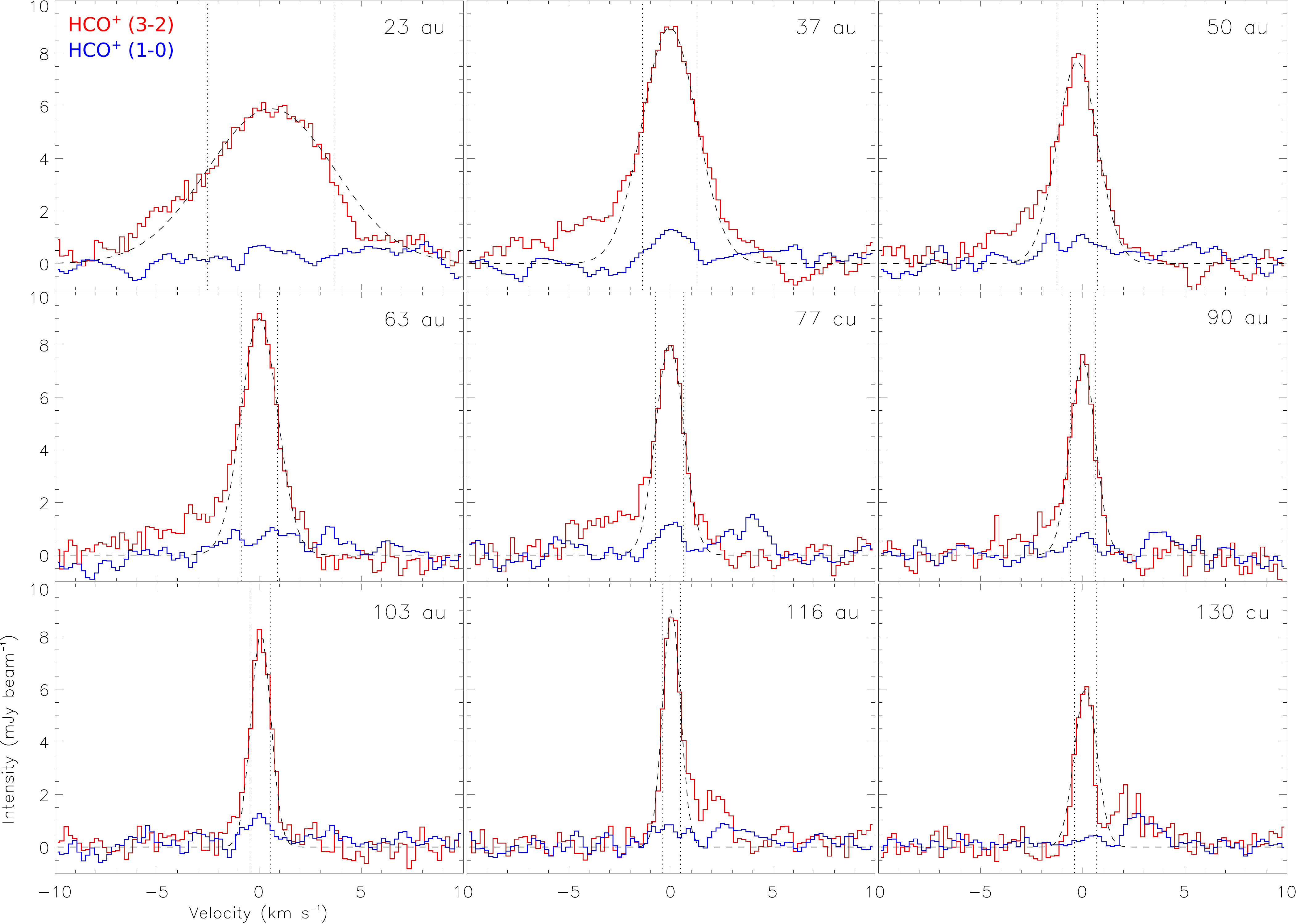}
\caption{Velocity-aligned stacked spectra of the HCO$^+$ (3--2; red) and (1--0; blue) emission of the HL Tau disk in different radial bins. The radius of the center of each radial bin is labeled at the upper right corner in each panel. The bin width is 7 au. Dashed curves present the Gaussian line fitting to the HCO$^+$ (3--2) spectra. In each panel, two dotted vertical lines denote the velocity range within the $\pm$1$\sigma$ Gaussian line width from the fitted line center, and this velocity range is adopted to compute the ratio of the HCO$^+$ (3--2) and (1--0) integrated intensities.}
\label{spec_r}
\end{figure*}

\subsection{HCO$^+$ density and temperature profiles}\label{TrNr}
We performed a joint analysis of the HCO$^+$ (3--2) and (1--0) and 1.1 mm and 2.9 mm continuum data to estimate the HCO$^+$ column density and the excitation temperature of the HCO$^+$ emission as a function of radius in the HL Tau disk.
Due to the proper motion of HL Tau \citep{Zacharias04}, 
 in our analysis, 
the coordinates of the 2.9 mm continuum and HCO$^+$ (1--0) data taken in 2014 are set to be relative offsets with respect to the continuum peak position measured in 2014 \citep{ALMA15}, 
and those of the 1.1 mm continuum and HCO$^+$ (3--2) data with respect to the peak position measured from our data taken in 2017 (Appendix~\ref{1.1mm_cont}).
We also compared the flux of the continuum emission at 1.1 mm measured with our observations and the previous measurements at 1.3 mm and 0.87 mm \citep{ALMA15}, and confirmed that our measured flux is consistent with the previous ALMA observations within the uncertainty of an absolution flux calibration of 5--10\% (Appendix~\ref{1.1mm_cont}).
In addition, the synthesized beams are different among these images, and it is coarsest in the HCO$^+$ (3--2) image cube.
We convolved the 1.1 mm and 2.9 mm continuum images and the HCO$^+$ (1--0) image cube to a beam size of $0\farcs1 \times 0\farcs09$ with a PA of 19$\degr$ the same as that of the HCO$^+$ (3--2) image cube.

\begin{figure}
\centering
\includegraphics[width=8cm]{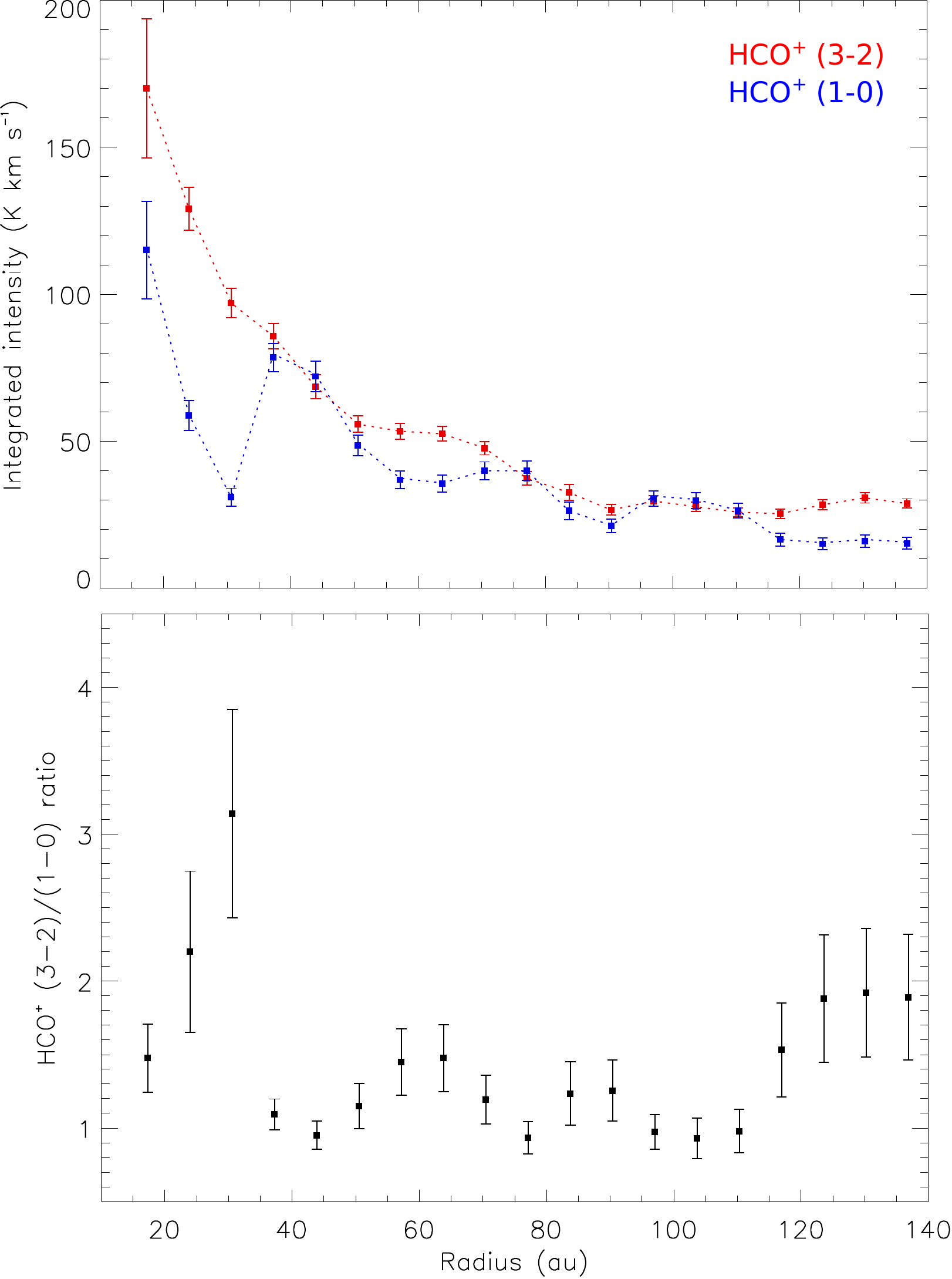}
\caption{Radial profiles of the integrated intensity of the HCO$^+$ (3--2; red) and (1--0; blue) emission (upper panel) and the (3--2) to (1--0) intensity ratio (lower panel) in the HL Tau disk. The HCO$^+$ (3--2) integrated intensity as a function of radius is measured from the velocity-aligned stacked spectra in radial bins. The HCO$^+$ (1--0) integrated intensity presented here is derived from the HCO$^+$ (3--2) integrated intensity and the measured (3--2) to (1--0) intensity ratio. The error bars are estimated from the error propagation of the observational noise. Dotted red and blue lines in the upper panel show the integrated intensity derived from Eq.~\ref{rt1} and \ref{rt2} with our estimated HCO$^+$ column density and excitation temperature.}
\label{hcop_ir}
\end{figure}

We first measured the intensity ratio of the HCO$^+$ (3--2) to (1--0) emission as a function of radius. 
To compare the two data sets, we re-gridded the HCO$^+$ (3--2) data cube to have the same velocity resolution and channels as the (1--0) data cube.
We find that the absorption in the HCO$^+$ (3--2) and (1--0) lines appears at different velocities and radii \citep{Yen16a}.  
To have a minimal effect of the different levels of absorption when comparing the integrated intensities of these two lines, 
we masked the absorption in their data cubes.
The absorption is identified from the azimuthally-averaged spectra without the alignment in radial bins, as described in Section \ref{MsVsys}. 
Joint masked velocity and radial ranges are adopted such that the absorption in both lines is all covered. 
Thus, the two data sets have the same masked regions.

Then we generated velocity-aligned stacked spectra in radial bins from the masked data of these two lines. 
Because the kinematics of the spiral structure is different from the observed Keplerian rotation in the disk (Section \ref{spiral}), 
we only selected data in the blueshifted part of the disk, corresponding to PA from 48$\degr$ to 228$\degr$, to avoid the spiral structure at the redshifted velocity.
The radius of the innermost bin is adopted to be the beam size of 0\farcs1 (14 au) because the velocity pattern in the innermost region within the one beam size cannot be well resolved, and the data cannot be properly aligned \citep{Yen16b}. 
The parameters for the alignment are adopted to be $M_\star$ of 2.14 $M_\sun$, $V_{\rm sys}$ of $V_{\rm LSR}$ = 7.14 km s$^{-1}$, the disk inclination angle of 47$\degr$, and the PA of the disk major axis of 138$\degr$.
In Fig.~\ref{spec_r}, we present the velocity-aligned stacked spectra of the HCO$^+$ (3--2) and (1--0) lines from their data cubes with the common masked regions. 
Every two radial bins are shown to reduce the number of panels. 

The S/N at the intensity peaks of the velocity-aligned stacked spectra of the HCO$^+$ (3--2) emission are above 20$\sigma$ in all radial bins.
For each radial bin, we fitted a Gaussian line profile to the velocity-aligned stacked spectrum of the HCO$^+$ (3--2) emission. 
At radii of 37--90 au, there is line wing emission in the HCO$^+$ (3--2) spectra at a relative velocity from $-2$ km s$^{-1}$ to $-5$ km s$^{-1}$ in addition to the central Gaussian-like component. 
Similar line wing emission is also seen in the HCO$^+$ (3--2) and (1--0) spectra at a relative velocity from 1.5 to 4.5 km s$^{-1}$ at radii larger than 110 au. 
Such an additional line wing is not seen in the velocity-aligned stacked spectra of more evolved sources, such as HD 163296 and  J16083070$-$3828268, where there is almost no ambient gas around the disks \citep{Yen16b, Yen18}.
These line wing features are likely due to the contamination from surrounding gas which is not in Keplerian rotation.

Compared to the HCO$^+$ (3--2) emission, the HCO$^+$ (1--0) emission is much fainter. 
The peak S/N of the velocity-aligned stacked spectra of the HCO$^+$ (1--0) emission are typically below 3$\sigma$--4$\sigma$.
Because of the low S/N of the HCO$^+$ (1--0) spectra, we did not fit their line profiles. 
In each radial bin, we integrated the velocity-aligned stacked spectra of the HCO$^+$ (3--2) and (1--0) emission over the channels within the 1$\sigma$ Gaussian line width from the centroid velocity, where the line width and the centroid velocity were from the Gaussian fitting to the HCO$^+$ (3--2) line.
The S/N of the integrated intensities of the HCO$^+$ (1--0) emission range from 4$\sigma$ to 10$\sigma$, 
and that of the HCO$^+$ (3--2) emission from 30$\sigma$ to 80$\sigma$. 
Then we computed the intensity ratio of the HCO$^+$ (3--2) to (1--0) emission as a function of radius from their integrated intensities.
These integrated intensities are measured from the same velocity and spatial ranges, where the emission is stronger, with the common masked regions for both lines. 
The measured radial profile of the intensity ratio of the HCO$^+$ (3--2) to (1--0) emission is shown in Fig.~\ref{hcop_ir}, 
which is computed after converting the integrated intensities in units of mJy~beam$^{-1}$~km s$^{-1}$ to the brightness temperature in  K~km~s$^{-1}$.

We adopted a similar procedure to measure the radial profile of the integrated intensity of the HCO$^+$ (3--2) emission.
We generated the velocity-aligned stacked spectra from the HCO$^+$ (3--2) data cube with its original velocity resolution of 0.1 km s$^{-1}$.
In addition, in this data cube, only regions showing the absorption in the HCO$^+$ (3--2) line were masked, 
different from the data cube used to measure the HCO$^+$ (3--2) to (1--0) intensity ratio, 
because a part of the HCO$^+$ (3--2) emission is in the common masked regions. 
The resultant spectra are similar to those in Fig.~\ref{spec_r}.
Then we fitted a Gaussian line profile to the velocity-aligned stacked spectrum of the HCO$^+$ (3--2) emission in each radial bin, 
and measured the peak intensity, line width, and integrated intensity.

Figure \ref{hcop_ir} presents our measured radial profile of the HCO$^+$ (3--2) integrated intensity in units of K km s$^{-1}$.
For comparison, we also plot the intensity profile of the HCO$^+$ (1--0) emission in Fig.~\ref{hcop_ir}. 
Because of the low S/N of the velocity-aligned stacked spectra of the HCO$^+$ (1--0) emission, 
it is uncertain to directly measure total integrated intensity of the HCO$^+$ (1--0) emission from the spectra. 
Thus, the intensity profile of the HCO$^+$ (1--0) emission shown here is derived by dividing the HCO$^+$ (3--2) integrated intensity by their intensity ratio.
In other words, we assume that the HCO$^+$ (3--2) and (1--0) emission lines trace the same volume in the HL Tau disk, 
and in each radial bin, the intensity ratio is a constant in the integrated velocity and spatial ranges. 
The radial profile of the HCO$^+$ (3--2) integrated intensity is relatively smooth. 
There are stronger variations in the radial profile of the (3--2) to (1--0) intensity ratio and thus the derived HCO$^+$ (1--0) integrated intensity.
In particular, spikes are seen at radii of $\sim$30 au, $\sim$65 au, and $\sim$130 au in the profile of the intensity ratio, 
corresponding to the dips in the profile of the HCO$^+$ (1--0) integrated intensity.

\begin{figure}
\centering
\includegraphics[width=8cm]{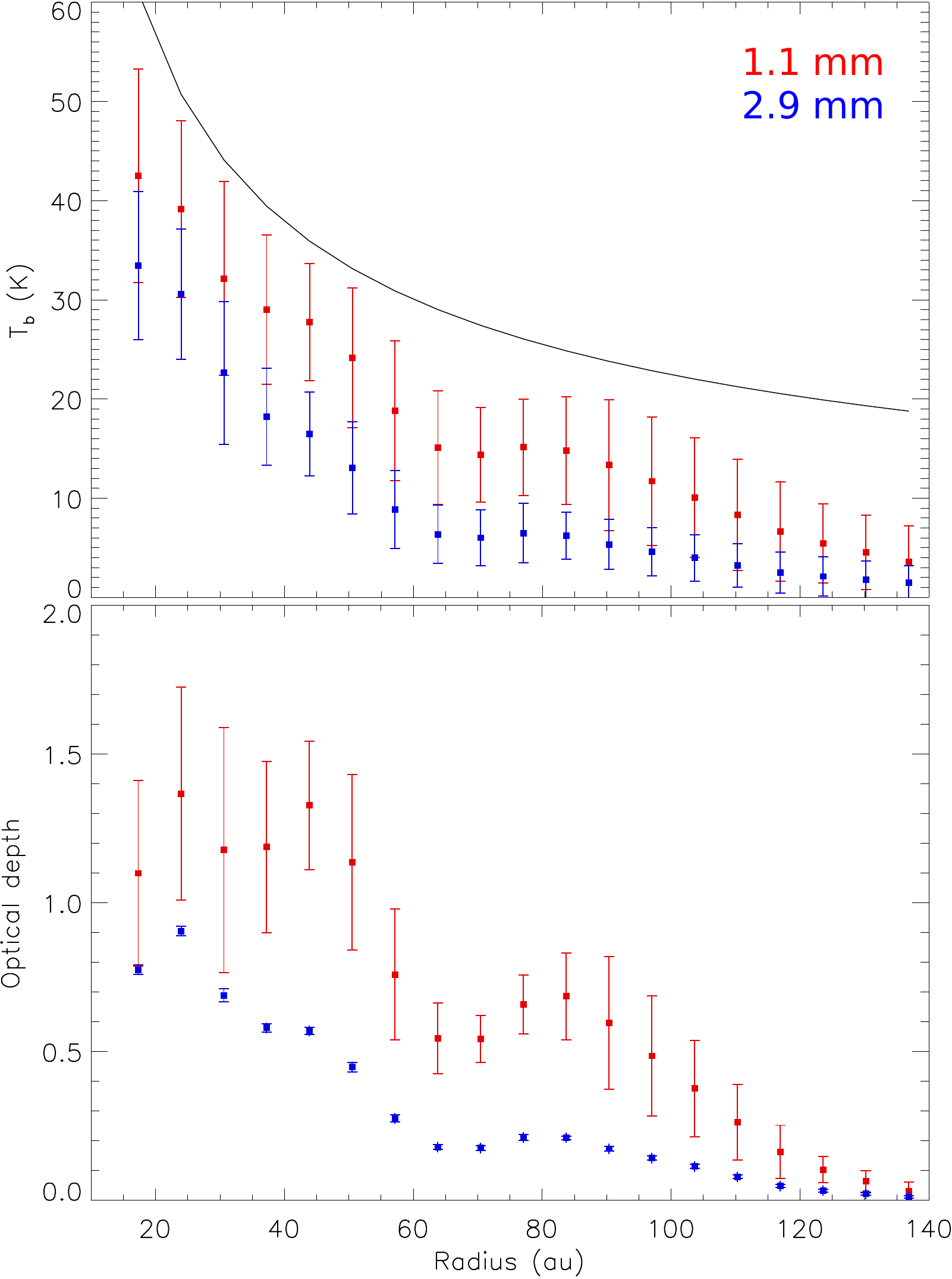}
\caption{Radial profiles of the brightness temperature of the 1.1 mm (red) and 2.9 mm (blue) continuum emission (upper panel) and their derived optical depths (lower panel) in the HL Tau disk. A solid curve in the upper panel presents the radial profile of the dust temperature adopted to derive the optical depth of the continuum emission.}
\label{cont_ir}
\end{figure}

Because the continuum emission at 1.1 mm and 2.9 mm is optically thick or marginally optically thick \citep{ALMA15, Pinte16}, 
the observed HCO$^+$ (3--2) and (1--0) intensities are likely suppressed due to the continuum opacity \citep[e.g.][]{Yen16a, Harsono18, Wu18}. 
To measure the continuum optical depths,
we first extracted the radial profiles of the 1.1 mm and 2.9 mm continuum intensities from the images. 
The profiles are azimuthally averaged over PA from 48$\degr$ to 228$\degr$, 
the same azimuthal range as that adopted to extract the line intensity.
The radial bin has a width of 0\farcs05, the same as that in the analysis of the lines.
Figure \ref{cont_ir} presents the measured radial profiles, 
and the intensity is converted to the brightness temperature.
We adopted the radial profile of the dust temperature from \citet{Okuzumi16}, $T_{\rm d} = 310 \times (R/1\ {\rm au})^{-0.57}$ K (solid curve in Fig.~\ref{cont_ir}). 
This dust temperature profile was derived by fitting a power-law profile to the brightness temperature profile of the disk around HL Tau at 0.87 mm \citep{Okuzumi16}, because the 0.87 mm continuum emission is optically thick \citep{ALMA15, Pinte16}.
Then we derived the radial profiles of the optical depths of the continuum emission ($\tau_{\rm d}$) with the radiative transfer equation as, 
\begin{equation}
\tau_{\rm d} = -\ln(1 - \frac{I_{\rm c, \nu}}{(B_{\nu}(T_{\rm d}) - B_{\nu}(T_{\rm bg}))}), 
\end{equation}
where $I_{\rm c, \nu}$ is the observed continuum intensity, $B_\nu(T)$ is the Planck function at temperature $T$, and $T_{\rm d}$ and $T_{\rm bg}$ are the dust temperature and the cosmic microwave background temperature of 2.73 K.
The derived radial profiles of the continuum optical depths are also shown in Fig.~\ref{cont_ir}.

\begin{figure}
\centering
\includegraphics[width=9cm]{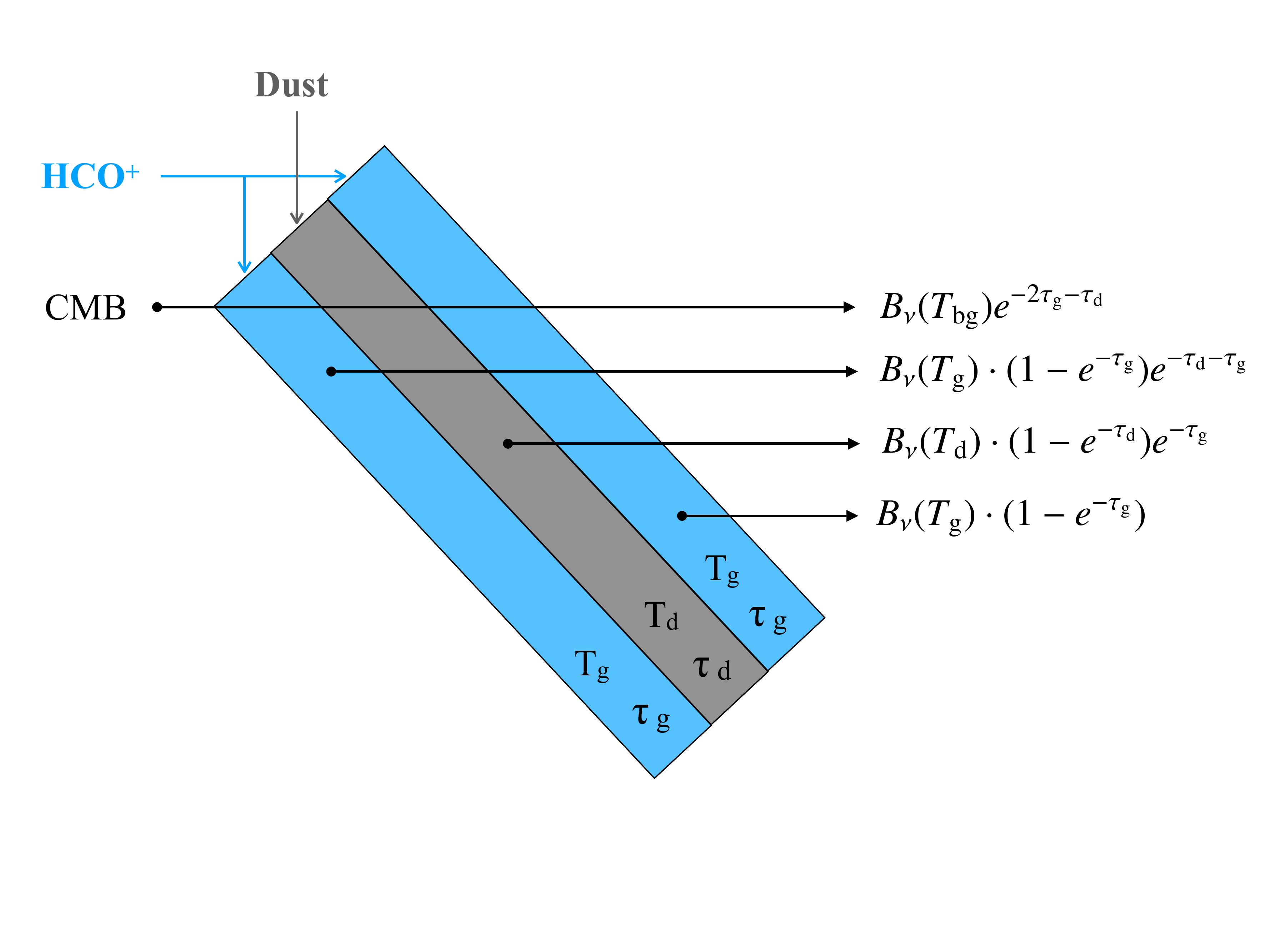}
\caption{Schematic figure of our simple three-layer model to correct the suppression of the continuum-subtracted line intensity by the continuum optical depth. In this model, dust and gas are totally segregated. The dust with a temperature of $T_{\rm d}$ and an optical depth of $\tau_{\rm d}$ is located in the mid-plane (grey color). The HCO$^+$ gas with an excitation temperature of $T_{\rm g}$ and an optical depth of $\tau_{\rm g}$ is located in an upper and lower layer (blue color). Contributions of the cosmic microwave background (CMB) and the three layers to the observed intensity are labelled on the right-hand side (Eq.~\ref{rt1}).}
\label{config}
\end{figure}

We adopt a simple three-layer model of the dust and HCO$^+$ gas distributions to correct the effect of the dust opacity on the line intensity (Fig.~\ref{config}).
We assume that the dust is concentrated in the disk mid-plane as a thin layer, and the HCO$^+$ gas is located above and below the layer of the dust. 
In other words, the dust and HCO$^+$ gas are totally segregated.
In this configuration, the HCO$^+$ emission from the far layer along the line of sight is first absorbed by the dust in the mid-plane and then by the HCO$^+$ gas in the near layer, and the dust continuum emission from the mid-plane is also absorbed by the HCO$^+$ gas in the near layer.
The HCO$^+$ emission from the near layer is not affected by the dust opacity.
We note that in case of a flared disk, the line of sight passes through various radial positions and heights, where the physical conditions could be different (see Fig.~\ref{hr}).
Nevertheless, because the actual height of the HCO$^+$ gas layer is not known, 
we adopt the geometrically thin approximation, 
and do not include any further correction for a light path passing through an inclined disk. 
Then, the observed intensities at the frequencies of the HCO$^+$ (3--2) and (1--0) lines ($I_{\nu}^\prime$) can be described as, 
\begin{eqnarray} \label{rt1}
I_{\nu}^\prime & = & B_\nu(T_{\rm bg})e^{-2\tau_{\rm g}-\tau_{\rm d}} \\ \nonumber
 & + & B_\nu(T_{\rm g})\cdot(1-e^{-\tau_{\rm g}})e^{-\tau_{\rm d}-\tau_{\rm g}} \\ \nonumber
 & + & B_\nu(T_{\rm d})\cdot(1-e^{-\tau_{\rm d}})e^{-\tau_{\rm g}} \\ \nonumber
 & + & B_\nu(T_{\rm g})\cdot(1-e^{-\tau_{\rm g}}) \\ \nonumber
 & - & B_\nu(T_{\rm bg}), 
\end{eqnarray}
where $\nu$ is the rest frequency of the HCO$^+$ (3--2) or (1--0) line, $T_{\rm g}$ is the excitation temperature ($T_{\rm ex}$) of HCO$^+$ (3--2) and (1--0), and $\tau_{\rm g}$ is the optical depth of the HCO$^+$ (3--2) or (1--0) emission. 
Here, we assume that $T_{\rm ex}$ of HCO$^+$ (3--2) and (1--0) are the same because non-LTE radiative calculations using RADEX \citep{radex07} show that HCO$^+$ is thermalized when the H$_2$ number density is higher than 10$^8$ cm$^{-3}$ at a temperature of 20--60 K, 
which are typical physical conditions in protoplanetary disks \citep{Williams11}.
Then the continuum-subtracted intensity of the HCO$^+$ emission ($I_{\nu}$) can be written as, 
\begin{equation}\label{rt2}
I_{\nu} = I_{\nu}^\prime - I_{\rm c, \nu},
\end{equation}
where $I_{\rm c, \nu}$ can be described as $[B_\nu(T_{\rm d})-B_\nu(T_{\rm bg})]\cdot(1-e^{-\tau_{\rm d}})$.

\begin{figure}
\centering
\includegraphics[width=8cm]{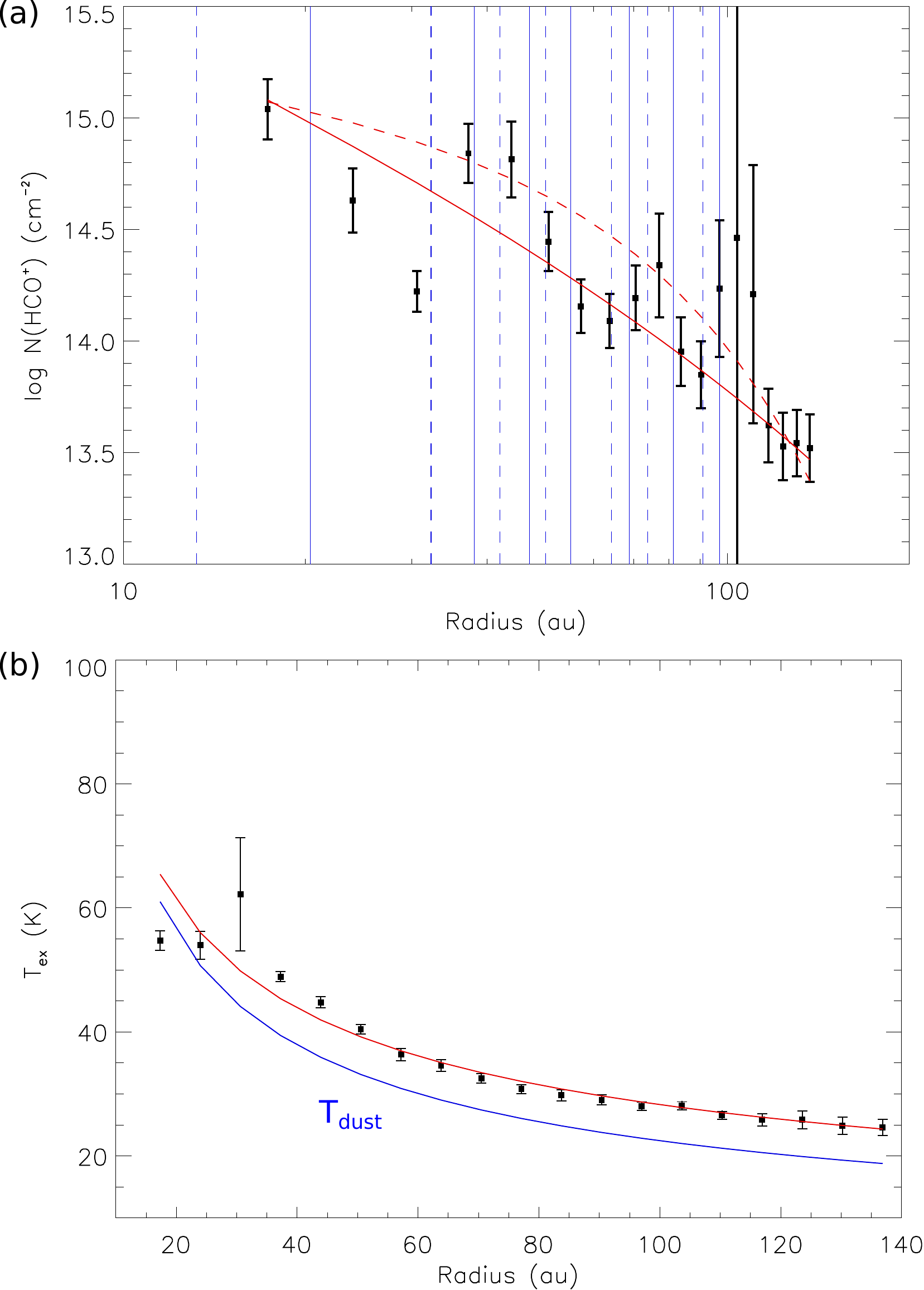}
\caption{(a) Radial profile of the column density of HCO$^+$. Vertical solid and dashed lines denote the radii of the rings and gaps in the HL Tau disk observed in the millimeter continuum emission \citep{ALMA15}, respectively. Red curves present the best-fit self-similar profiles to the observed data points. The red solid curve is the fit by excluding the data points at radii between 20 au and 35 au, and the red dashed curve is the fit by further excluding the data points at radii between 45 au and 100 au. (b) Radial profile of the excitation temperature of the HCO$^+$ (3--2) and (1--0) lines. A blue curve shows the dust temperature profile from \citet{Okuzumi16}. A red curve presents the power-law profile fitted to the data points, $256\cdot(R/{\rm 1\ au})^{-0.48}$ K.}
\label{Nhcop}
\end{figure}

We solved Eq.~\ref{rt1} and \ref{rt2} for the HCO$^+$ (3--2) and (1--0) lines to derive the column density and $T_{\rm ex}$ of HCO$^+$ for each radial bin. 
The inputs are the peak intensities and line widths of the velocity-aligned stacked spectra of the HCO$^+$ (3--2) emission, the intensity ratios of the HCO$^+$ (3--2) to (1--0) emission (Fig.~\ref{hcop_ir}), the dust temperature profile from \citet{Okuzumi16}, and the derived continuum optical depths at 1.1 mm and 2.9 mm (Fig.~\ref{cont_ir}). 
For each radial bin, there are two unknowns, the column density and $T_{\rm ex}$ of HCO$^+$, less than the number of the inputs.
Thus, Eq.~\ref{rt1} and \ref{rt2} can be solved.
For a given set of the column density, $T_{\rm ex}$, and the line width, 
we first compute $\tau_{\rm g}$ from these parameters.
The calculation of $\tau_{\rm g}$ is explained in detail in \citet{Mangum15}.
Then, we plug the computed $\tau_{\rm g}$ together with the dust temperature and the continuum optical depths into Eq.~\ref{rt1} and \ref{rt2}, 
derive the intensities of the HCO$^+$ (3--2) and (1--0) lines for this given set of the column density and $T_{\rm ex}$, 
and compare with the observed values. 
To automatically solve Eq.~\ref{rt1} and \ref{rt2} with this procedure,
we used the IDL routine {\it MPFIT} \citep{Markwardt09} for the minimization to find the column density and $T_{\rm ex}$ that can reproduce the observed HCO$^+$ (3--2) and (1--0) intensities, 
For all radial bins, the solutions, which exactly reproduce the observed intensities, were successfully obtained (dotted lines in Fig.~\ref{hcop_ir}).

The radial profiles of the column density and $T_{\rm ex}$ of HCO$^+$ estimated from this analysis are shown in Fig.~\ref{Nhcop}.
A gap at a radius of 30 au is clearly seen, corresponding to the radius of the highest spike in the radial profile of the HCO$^+$ (3--2) to (1--0) intensity ratio and the lowest dip in the HCO$^+$ (1--0) intensity profile (Fig.~\ref{hcop_ir}). 
The estimated column density at a radius of 30 au is $(1.7\pm0.4) \times 10^{14}$ cm$^{-2}$.
By interpolating the estimated column densities at radii of 17 au and 37 au with a power-law function, 
the column density is derived to be $7.8 \times 10^{14}$ cm$^{-2}$ at a radius of 30 au if there is no gap. 
Assuming that the HCO$^+$ abundance is constant at radii between 17 au and 37 au, 
the depth of this HCO$^+$ gas gap is estimated to be 4--8,
where the depth is defined as the ratio between the estimated column density in the gap and the column density derived from the interpolation from the neighboring radii.
There is possibly a second gap at a radius of 60--70 au, 
but the contrast in the column density between the dip and the neighboring data points is not high compared to the error bars. 

We fitted the column density ($\Sigma$) profile with a self-similar solution for an accretion disk \citep{Lynden74, Hartmann88}, 
\begin{equation}
\Sigma (r) = \Sigma_0 (\frac{r}{R_{\rm c}})^{-\gamma} \exp[-(\frac{r}{R_{\rm c}})^{(2-\gamma)}].
\end{equation}
Excluding the data points at radii between 20 au and 35 au, which belong to the inner gap, 
the best-fit profile is $\Sigma_0 = 2.5 \times 10^{14}$ cm$^{-2}$, $R_{\rm c} = 82$ au, and $\gamma = 1.2$ (red solid curve in Fig.~\ref{Nhcop}a).
This profile is comparable to that measured from the ALMA continuum data \citep{Akiyama16} and is slightly steeper than the typical profiles with $\gamma < 1$ in protoplanetary disks measured from the millimeter continuum emission \citep{Andrews09, Andrews10, Tazzari17}.
When we further exclude the data points at radii between 45 au and 100 au, where there is possibly a second gap, 
the best-fit profile becomes $\Sigma_0 = 7.6 \times 10^{14}$ cm$^{-2}$, $R_{\rm c} = 66$ au, and $\gamma = 0.4$ (red dashed curve).
This $\gamma$ is in the typical range in protoplanetary disks. 
The data points at radii close to 60 au are indeed lower than the overall fitted profile, 
but the difference is less than 2$\sigma$. 
By integrating the column density profile, 
the gas mass of the HL Tau disk is estimated to be $4.4 \times 10^{-12}/X({\rm HCO^+})$ $M_\sun$, 
where $X({\rm HCO^+})$ is the HCO$^+$ abundance relative to H$_2$, 
assuming a constant $X({\rm HCO^+})$ in the HL Tau disk.

\begin{figure}
\centering
\includegraphics[width=8cm]{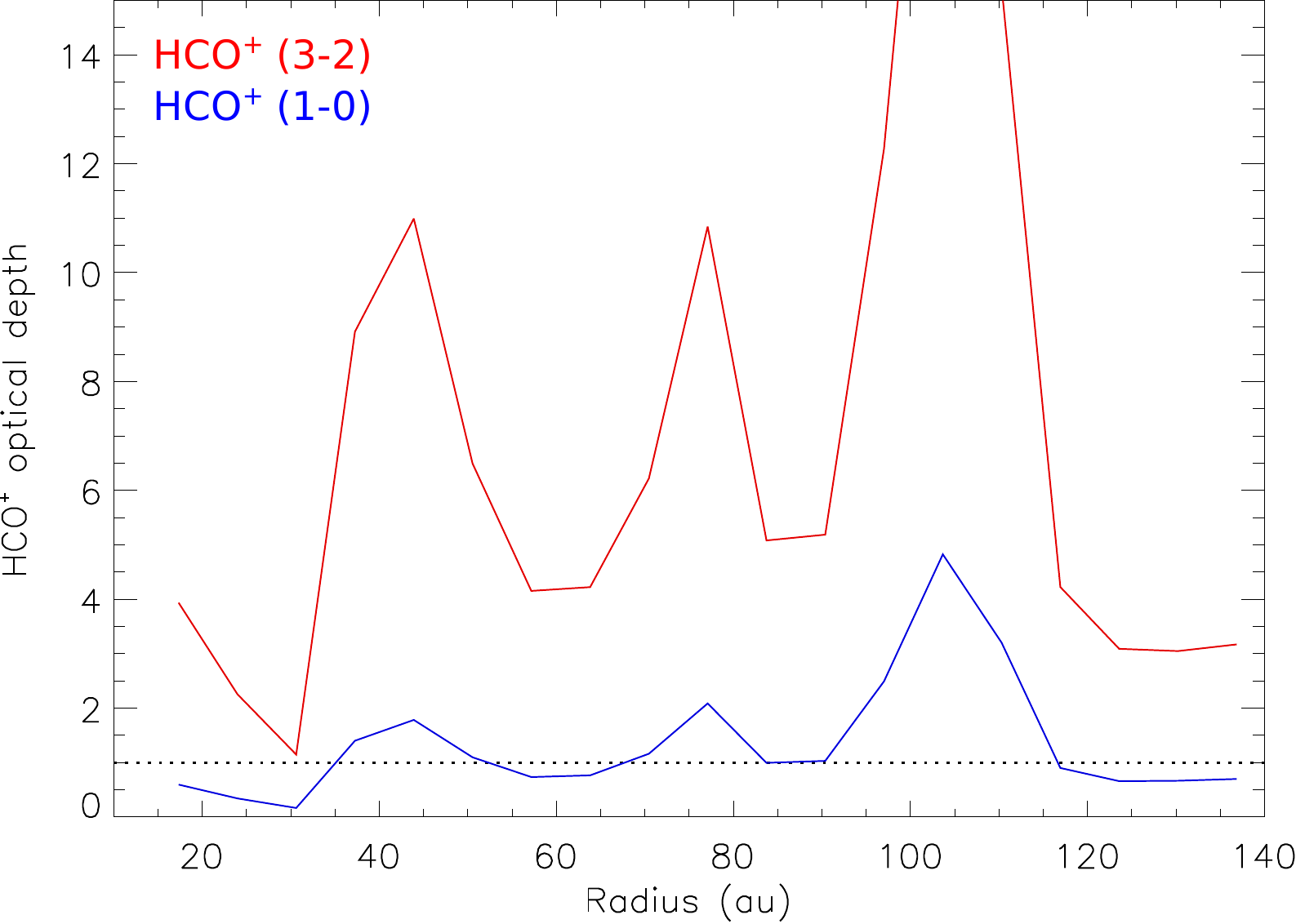}
\caption{Radial profiles of the optical depths of the HCO$^+$ (3--2; red) and (1--0; blue) lines at the peak velocity. The horizontal dotted line denotes an optical depth of one.}
\label{hcop_tau}
\end{figure}

The overall $T_{\rm ex}$ profile can be described by a power-law profile of $256\cdot(R/{\rm 1\ au})^{-0.48}$ K for $15 < R < 140$ au (red curve in Fig.~\ref{Nhcop}b).
The estimated $T_{\rm ex}$ is higher in the HCO$^+$ gas gap. 
Nevertheless, the error bar is also large at the radius of the HCO$^+$ gas gap because of the lower S/N.
Generally, $T_{\rm ex}$ of the HCO$^+$ lines is higher than the dust temperature by 5--10 K. 
On the other hand, the estimated $T_{\rm ex}$ of the HCO$^+$ lines is lower than the brightness temperature of the CO (1--0) emission measured with the ALMA observations at an angular resolution of 0\farcs07 by a factor of two to three at all radii. 
The radial profile of the optical depths of the HCO$^+$ lines at the peak velocity derived from this analysis are presented in Fig.~\ref{hcop_tau}. 
The analysis suggests that the HCO$^+$ (3--2) line is optically thick with $\tau_{\rm g}$ ranging from 2.5 to $>$10 at the peak velocity, except for that at the radius of the inner gap, where $\tau_{\rm g}$ is close to unity. 
The HCO$^+$ (1--0) line is optically thin or marginally optically thick with $\tau_{\rm g}$ typically smaller than 3.
The uncertainty in the HCO$^+$ column density is large at a radius of $\sim$100 au, and thus the estimated optical depths of the HCO$^+$ (3--2) and (1--0) lines are also very uncertain at that radius.

\begin{figure}
\centering
\includegraphics[width=8cm]{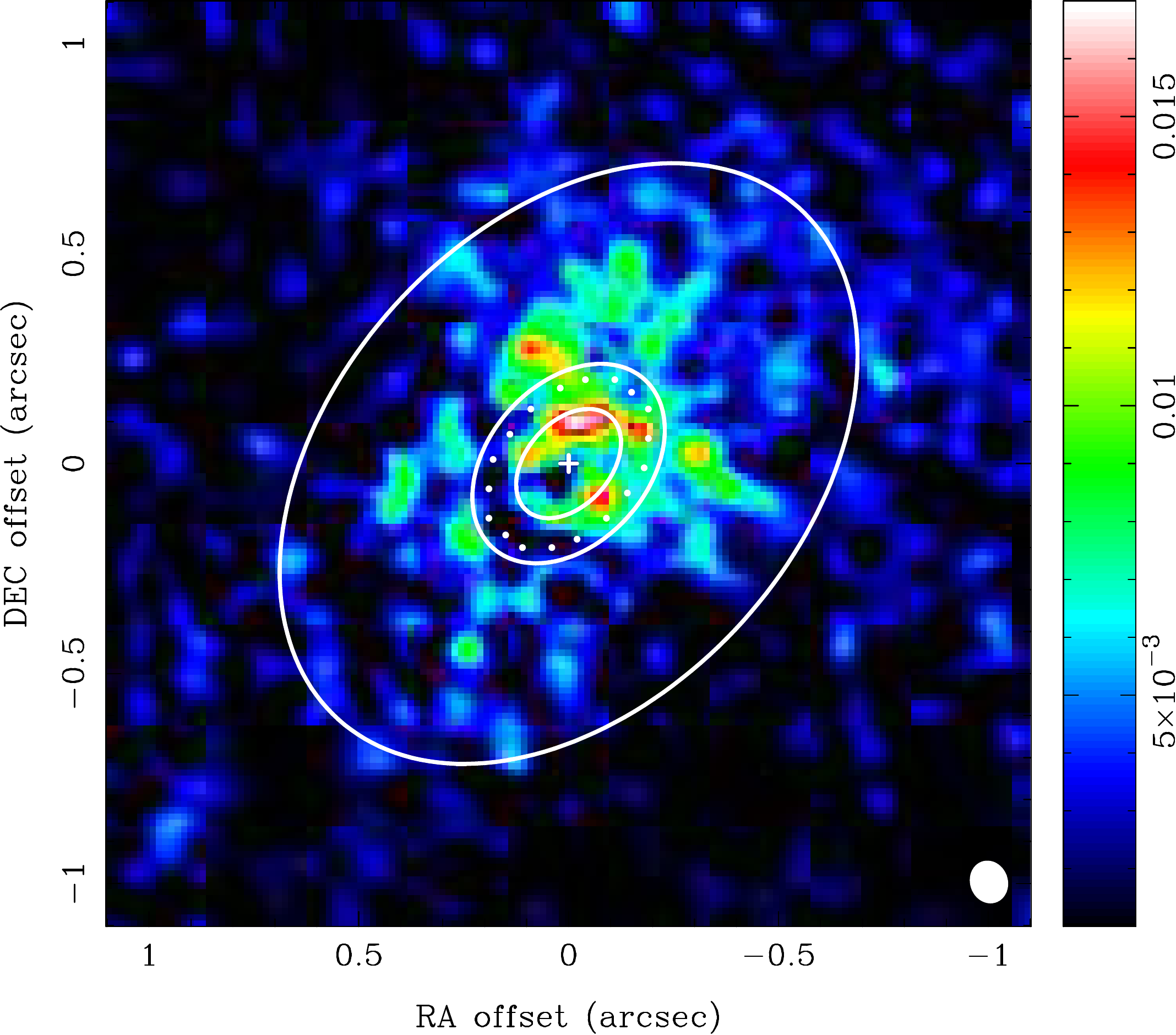}
\caption{Moment 0 map of the HCO$^+$ (1--0) emission in the HL Tau disk generated with the Keplerian masking technique. Open ellipses from inside to outside show the first and second dust rings at radii of 20.4 au and 38.1 au observed by \citet{ALMA15} and the disk radius of 115 au observed in the 1.1 mm continuum emission, respectively. Dots delineate the dust gap at a radius of 32.3 au observed by \citet{ALMA15}. A cross denote the position of HL Tau. A filled white ellipse presents the beam size.}
\label{hcop10_mom0}
\end{figure}

The HCO$^+$ gas gap at a radius of 30 au cannot be directly seen in the moment 0 and 8 maps of the HCO$^+$ (3--2) emission (Fig.~\ref{hcop_mom} and \ref{hcop_mom1}), 
although the HCO$^+$ (3--2) emission is almost optically thin with the optical depth of $\sim$1 in the gap.
This is due to the higher $T_{\rm ex}$ in the gap, and thus the HCO$^+$ (3--2) intensities in the gap and at the adjacent radii are comparable. 
The HCO$^+$ (1--0) emission is optically thiner, 
and the contrast between the HCO$^+$ (1--0) intensities in the gap and at the adjacent radii is expected to be higher.
To show the intensity distribution of the HCO$^+$ (1--0) emission, 
we adopted the Keplerian masking technique \citep[e.g.,][]{Salinas17, Ansdell18}, generated a mask using the Keplerian rotational velocity pattern and the radial profile of the line widths measured from the HCO$^+$ (3--2) emission, and made a moment 0 map of the HCO$^+$ (1--0) emission (Fig.~\ref{hcop10_mom0}).
The moment 0 map shows that the HCO$^+$ (1--0) emission is bright in the northwest in the disk, similar to the HCO$^+$ (3--2) emission.
A deficit of the HCO$^+$ (1--0) intensity is seen at a radius of $\sim$30 au in the southeastern side of the disk, 
and the HCO$^+$ (1--0) emission is more clearly detected at 3--5$\sigma$ at outer radii of $\sim$40--60 au. 
Thus, the HCO$^+$ (1--0) moment 0 map shows signs of a gas gap in the southeastern side of the disk, 
as in the azimuthally averaged column density profile derived from the LTE analysis of the HCO$^+$ (3--2) and (1--0) lines.
The northwestern side of the disk does not show any clear deficit of the HCO$^+$ (1--0) intensity at a radius of 30 au, which could be due to the contamination from the spiral structure in the west.

We note that the overall column density scale of HCO$^+$ estimated in this work is an order of magnitude lower than in \citet{Yen16a}, which is on the order of 10$^{14}$--10$^{15}$ cm$^{-2}$.
This difference is caused by different $T_{\rm ex}$ adopted in the analyses. 
In this work, $T_{\rm ex}$ of the HCO$^+$ (3--2) and (1--0) lines is derived from the line ratio. 
In \citet{Yen16a}, $T_{\rm ex}$ is adopted to be the brightness temperature of the CO (1--0) emission observed with ALMA because only the HCO$^+$ (1--0) data were available. 
Thus, $T_{\rm ex}$ in this work is a factor of two to three lower than in \citet{Yen16a}. 
When $T_{\rm ex}$ is much higher than the upper energy level of the HCO$^+$ (1--0) line of 4.28 K, 
the column density derived from the integrated intensity of the HCO$^+$ (1--0) emission is approximately proportional to ${T_{\rm ex}}^2$ \citep{Mangum15}.
As a result, our estimated HCO$^+$ column density is lower than that in \citet{Yen16a}. 
On the other hand, the estimated depth of the gap at a radius of 30 au is a factor of two to three deeper than that in \citep{Yen16a}.
This is likely due to the better correction of the line suppression due to the continuum optical depth, the better estimation of $T_{\rm ex}$ of the HCO$^+$ lines, and the well resolved disk rotation in this work.

We also note that the column density, $T_{\rm ex}$, and $\tau_{\rm g}$ are derived with the line width measured from the velocity-aligned stacked spectra of the HCO$^+$ (3--2) emission.
Our method of the alignment can remove the contribution to the line width of a velocity-aligned stacked spectrum caused by different motional velocities in an averaged area \citep{Yen16b}.
Nevertheless, this removal is limited by the finite angular resolution. 
The velocity pattern within one beam size cannot be well resolved, 
and spectra extracted from an area smaller than the beam size cannot be properly aligned, 
leading to line broadening.  
In the case of our observations of HL Tau with a beam size of 0\farcs1, 
along the disk major axis, 
the difference in the rotational velocity across one beam size is $\sim$5 km s$^{-1}$ at a radius of 0\farcs1 and is $\sim$0.4 km s$^{-1}$ at a radius of 0\farcs5.
The velocity difference within one beam size also decreases when the averaged area is located closer to the minor axis due to the projection onto the line of sight.
Thus, the broad line width of the velocity-aligned stacked spectrum at the smallest radius, 23 au in Fig.~\ref{spec_r}, is likely caused by line broadening due to a not well resolved velocity pattern within the beam size, 
while at outer radii, the measured line widths of the velocity-aligned stacked spectra are expected to be dominated by the thermal and turbulent line widths.
Nevertheless, the measured integrated intensity is robust because this method of the velocity alignment conserves the flux \citep{Yen16b}. 
Due to the overestimated line width at the inner radius, 
the column density, $T_{\rm ex}$, and $\tau_{\rm g}$ at the inner radius can be underestimated. 

\begin{figure}
\centering
\includegraphics[width=8cm]{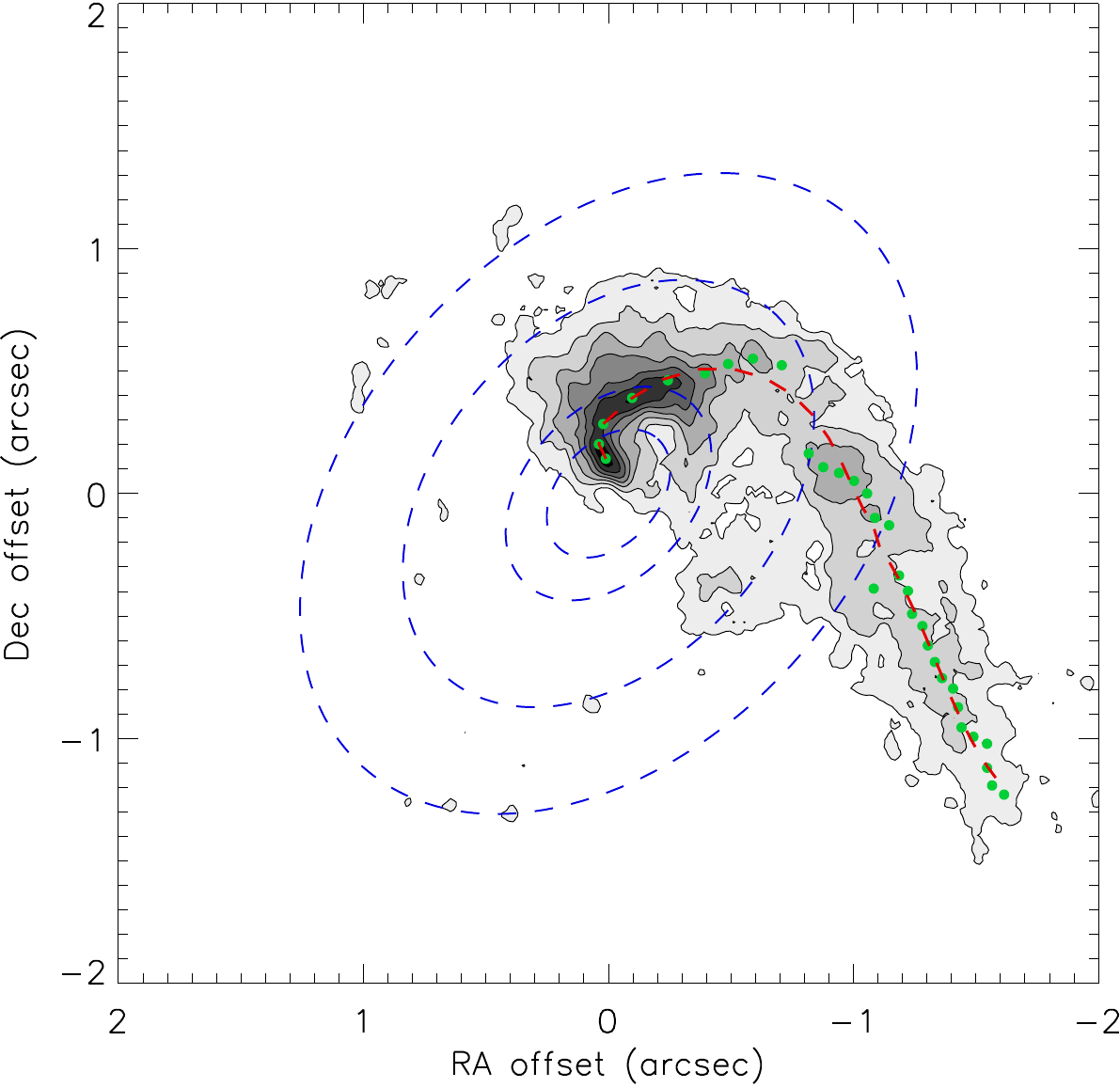}
\caption{Moment 0 map of the HCO$^+$ (3--2) emission integrated from $V_{\rm LSR}$ of 7.3 km s$^{-1}$ to 11.2 km s$^{-1}$, where the spiral structure is seen in the velocity channel maps. Zero offset refers to the position of HL Tau. Blue dashed ellipses denote the de-projected radii of 0\farcs3, 0\farcs5, 1$\arcsec$, and 1\farcs5. Green dots present the measured peak positions as a function of de-projected radius. A red dashed curve presents the fitted polynomials to the green dots. Contour levels start from 5$\sigma$ in steps of 5$\sigma$, where 1$\sigma$ is 1.6 mJy beam$^{-1}$ km s$^{-1}$.}
\label{spiral_mom0}
\end{figure}

\subsection{kinematics of the one-arm spiral}\label{spiral}
\begin{figure}
\centering
\includegraphics[width=8.5cm]{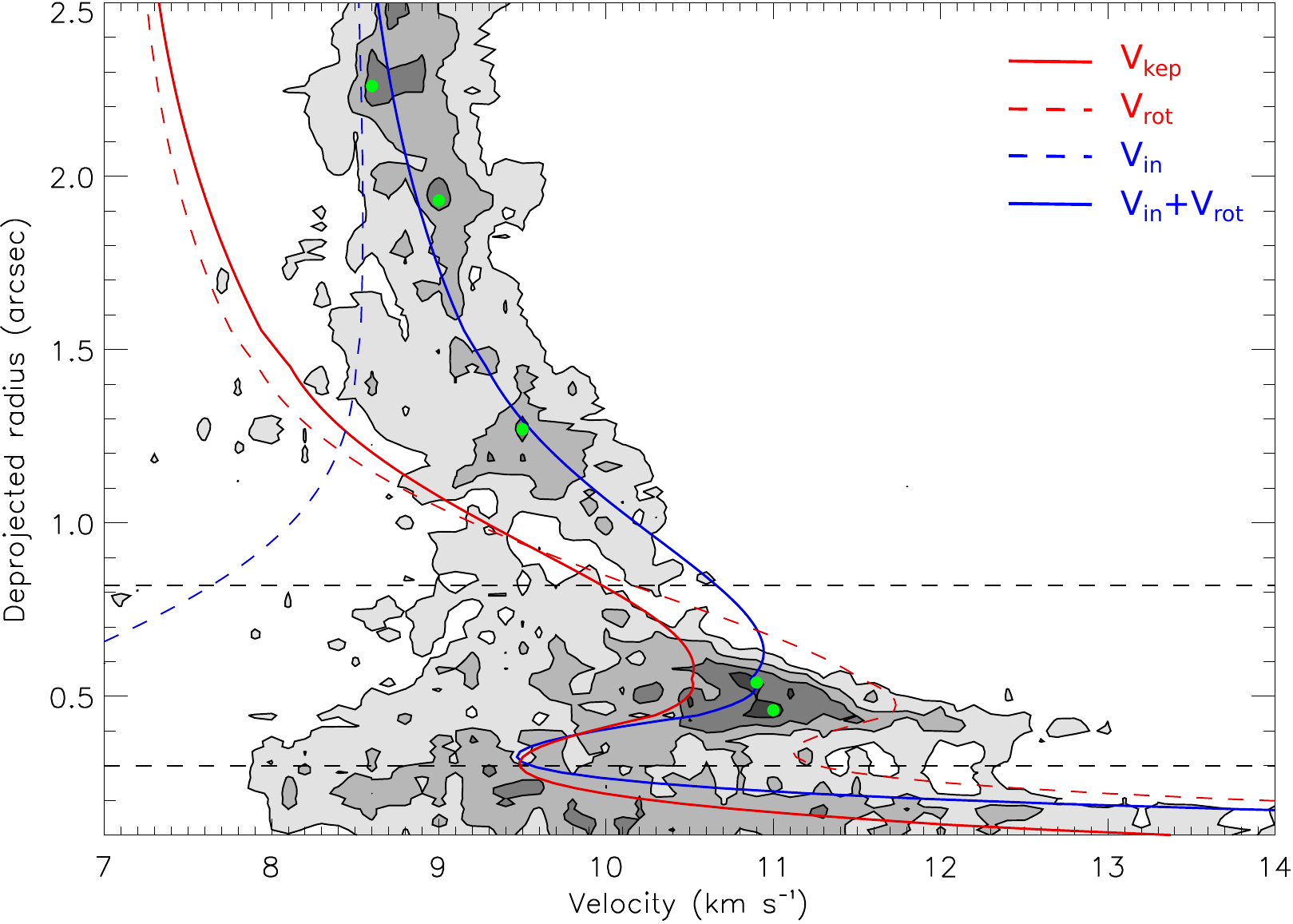}
\caption{Position--velocity diagram along the spiral structure (red dashed curve in Fig.~\ref{spiral_mom0}). Upper and lower horizontal dashed lines denote radii of 0\farcs8 and 0\farcs3, repsectively. A red solid curve presents the velocity of the Keplerian rotation projected onto the line of sight. A blue solid curve shows the line-of-sight velocity of the infalling+rotational motion that can describe the observed velocity profile along the spiral structure. The separated contributions to the line-of-sight velocity by the infalling and rotational motions are presented by blue and red dashed curves, respectively. Five green dots denote the representative positions (the local peaks) in the P--V diagram that we selected to derive the velocity profiles of the infalling and rotational motions. Contour levels start from 3$\sigma$ in steps of 3$\sigma$, where 1$\sigma$ is 2.5 mJy beam$^{-1}$.}
\label{spiral_pv}
\end{figure}

Figure \ref{spiral_mom0} presents the moment 0 map of the HCO$^+$ (3--2) emission integrated from $V_{\rm LSR}$ of 7.3 km s$^{-1}$ to 11.2 km s$^{-1}$, where the spiral structure is seen in the velocity channel maps.
To describe the morphology of the spiral structure, we measured the peak positions as a function of de-projected radius in steps of 0\farcs1. 
For simplicity, we assume that the spiral structure is in the mid-plane of the disk because its actual three-dimensional (3D) orientation is not known. 
For each de-projected radius, we extracted an intensity profile along the azimuthal direction from the moment 0 map and measured its peak position, which is labelled as a green dot in Fig.~\ref{spiral_mom0}.
In our analysis, the positions of these data points are described in polar coordinates. 
We found that a single spiral function or polynomial cannot fit the measured peak positions. 
Thus, we separated the spiral structure into three regions, $r < 0\farcs5$, $0\farcs5 < r < 1\farcs5$, and $r > 1\farcs5$, because of the changes in the curvature. 
We fitted the innermost region, where the curvature is large, with a third degree polynomial, 
and the outer two regions with second degree polynomials. 
The parameters of these polynomials are listed in Table \ref{poly}.
We linked these three polynomials to delineate the spiral structure, shown as a red dashed curve in Fig.~\ref{spiral_mom0}. 
The exact parameters of these polynomials are not essential. 
Here we only used this red dashed curve to extract a P--V diagram along the spiral structure for analysis of its velocity structure.
The P--V diagram along the spiral structure is presented in Fig.~\ref{spiral_pv}.

\begin{table}
\caption{Polynomials to fit the spiral}\label{poly}
\centering
\begin{tabular}{ccccc}
\hline\hline
\multicolumn{5}{c}{$\theta = a_0 + a_1 r + a_2 r^2 + a_3 r^3$} \\
 \hline
Radial range & $a_0$ & $a_1$ & $a_2$ &  $a_3$ \\
\hline\hline
$<$0\farcs5 & 285 & 328 & $-$484 & $-$295 \\
0\farcs5--1\farcs5 & 408 & -255 & 80 & 0 \\
$>$1\farcs5 & 256 & -47 & 7 & 0 \\
\hline\hline
\end{tabular}
\tablecomments{$r$ and $\theta$ are the de-projected radius and azimuthal angle, respectively.}
\end{table}

We compared the observed velocity profile along the spiral structure with the Keplerian rotation of the HL Tau disk (red solid curve in Fig.~\ref{spiral_pv}). 
At de-projected radii smaller than 0\farcs3 (lower horizontal dashed line), 
the observed velocity profile more or less follows the Keplerian rotation. 
This radius of 0\farcs3 is close to the location of the brightest peak in the moment 0 map of the HCO$^+$ (3--2) emission in Fig.~\ref{hcop_mom}.
On the other hand, 
at radii between 0\farcs4 and 0\farcs6, the intensity peak is located at a velocity more redshifted than the Keplerian velocity by 0.5 km s$^{-1}$. 
Outside the HL Tau disk with an outer radius of 0\farcs82 in the continuum (upper horizontal dashed line), 
the emission is also distributed at velocities more redshifted than the Keplerian velocities by 1--1.5 km s$^{-1}$, 
and the velocity excess increases with increasing radii. 

The spiral structure is more redshifted than the expectation from the Keplerian rotation at radii larger than 0\farcs4, 
and it is extended to the southwest of HL Tau, where is the near side of the disk and the envelope around HL Tau.
Thus, the observed velocity excess could be explained if the spiral structure is infalling toward the HL Tau disk because the infalling motion in the near side of the disk and the envelope induces a redshifted velocity.

To search for a velocity profile of the infalling and rotational motions that can explain the observed velocity pattern along the spiral structure, 
we first selected representative data points from the P--V diagram, shown as green dots in Fig.~\ref{spiral_pv}.
These data points correspond to the velocities and de-projected radii of the local peaks in the P--V diagram. 
The radial profiles of the infalling ($V_{\rm in}$) and rotational ($V_{\rm rot}$) velocities of the spiral structure are assumed to be power-law function as, 
\begin{eqnarray}
V_{\rm in}(r) = V_{\rm in,0} \cdot (r/1\arcsec)^p, \\
V_{\rm rot}(r) = V_{\rm rot,0} \cdot (r/1\arcsec)^q. 
\end{eqnarray}
Then the line-of-sight velocity ($V_{\rm los}$) of the infalling and rotational motion can be computed with 
\begin{equation}\label{vspiral}
V_{\rm los}(r) = V_{\rm in}(r) \cdot \sin i \cos \theta - V_{\rm rot}(r) \cdot \sin i \sin \theta,
\end{equation}
where $i$ is the inclination angle of the disk and $\theta$ is the angle between the radial direction and the minor axis.
We assume that the rotation of the spiral structure has a conserved angular momentum, meaning that $q = -1$.
We fitted Eq.~\ref{vspiral} to the representative data points from the P--V diagram, 
and we obtained $V_{\rm in,0}$ of $-2.2$ km s$^{-1}$, $p$ of $-0.16$, and $V_{\rm rot,0}$ of 3.6 km s$^{-1}$.
$V_{\rm in,0}$ is negative because the infalling motion is toward the center. 
A blue solid curve in Fig.~\ref{spiral_pv} presents the derived profile of $V_{\rm los}$ of the infalling and rotational motion.
This profile indeed can explain the overall observed velocity profile along the spiral structure.
Nevertheless, we note that in the P--V diagram, there is a lack of emission at the derived $V_{\rm los}$ at a radius of 0\farcs8, which is the disk radius in the continuum. 

\begin{figure*}
\centering
\includegraphics[width=16cm]{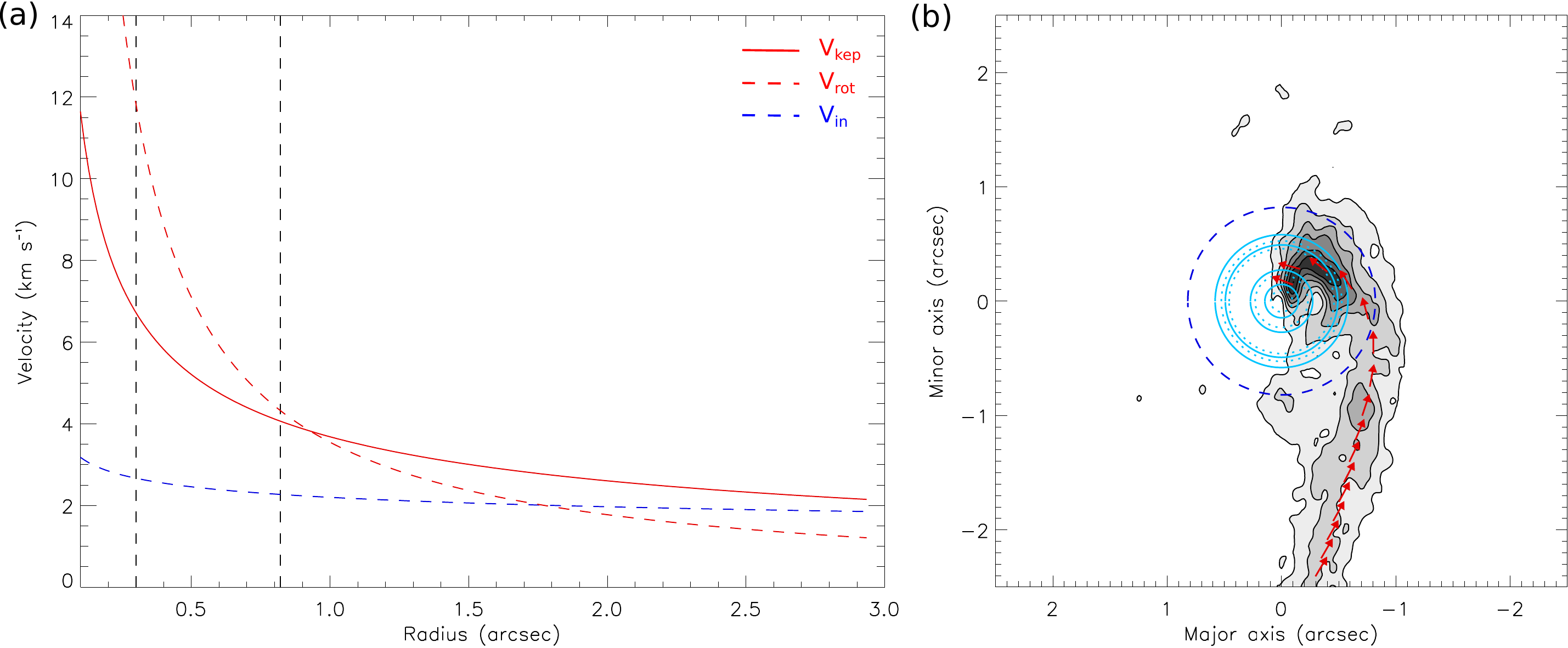}
\caption{(a) Radial profiles of the infalling (blue dashed) and rotational (red dashed) velocities, which can explain the observed velocity profile along the spiral structure, in comparison with the velocity profile of the Keplerian rotation of the HL Tau disk (red solid). An inner vertical dashed line denotes the radius of 0\farcs3, where the observed velocity profile starts to follow the Keplerian rotation, and the outer one denotes the disk radius of 0\farcs82 observed in the continuum emission. (b) De-projection of the moment 0 map of the redshifted HCO$^+$ (3--2) emission in Fig.~\ref{spiral_mom0} on the assumption that the spiral structure is in the disk mid-plane. A dark blue dashed circle delineates the disk size observed in the continuum emission. Light blue solid and dotted circles denote the radii of the dust rings and gaps observed in the continuum emission, respectively \citep{ALMA15}. Red arrows present the direction of the motional velocity along the spiral structure, computed from the ratio between the derived profiles of the infalling and rotational velocities in (a). Contour levels start from 5$\sigma$ in steps of 5$\sigma$, where 1$\sigma$ is 1.6 mJy beam$^{-1}$ km s$^{-1}$.}
\label{vprofile}
\end{figure*}

Figure \ref{vprofile}a compares the radial profiles of the derived infalling and rotational velocities with the Keplerian rotation of the HL Tau disk. 
The derived infalling velocity is slower than the free-fall velocity toward a central stellar mass of 2.1 $M_\sun$, 
and its profile is also shallower than $r^{-0.5}$.
The derived rotational motion dominates over the infalling motion at radii smaller than 1\farcs7, 
and exceeds the Keplerian velocity at a radius of 0\farcs9, close to the disk radius observed in the continuum emission.
In Fig.~\ref{vprofile}b, we de-project the moment 0 map of the spiral structure on the assumption that it is in the disk mid-plane in comparison with the radii of the dust rings and gaps observed in the continuum emission \citep{ALMA15}.
In addition, we plot the direction of the motional velocity by combining the derived infalling and rotational velocities. 
At outer radii, the direction of the motion approximately follows the tangential direction of the observed spiral structure. 
At radii smaller than 0\farcs5, where the derived rotational velocity is higher than the infalling velocity by a factor of two, 
the direction of the motion is more along the azimuthal direction.

We note that the profiles of the infalling and rotational velocities derived here are based on the assumption that the spiral structure is in the disk mid-plane and that its angular momentum is conserved. 
If the spiral structure is not in the disk mid-plane or the angular momentum is not conserved, 
the derived profiles of the infalling and rotational velocities are not valid. 
Nevertheless, the qualitative trends as the following are valid regardless of the actual 3D orientation, (1) the spiral structure is infalling toward the disk and (2) its rotational motion dominates over its infalling motion at inner radii.  
This is because the spiral structure at radii larger than 0\farcs6 is located on the near side and is more redshifted than the Keplerian rotation, suggesting the presence of the infalling motion. 
In addition, at radii smaller than 0\farcs6, the spiral structure is located in the northwest of HL Tau, where the far side is.
The infalling and rotational motions there induce blueshifted and redshifted velocities, respectively.  
The observed emission in the spiral structure in the northwest is still redshifted. 
Thus, at the inner radii, the rotational motion needs to dominate over the infalling motion.

\section{Discussion}\label{discuss}
\subsection{Formation of the HCO$^+$ gas gap}\label{discuss1}
\subsubsection{Gap opened by a planet?}
The radius of the observed HCO$^+$ gas gap is coincident with the gap D2 at a radius of 32 au in the millimeter continuum emission detected with ALMA \citep{ALMA15}. 
The depth of the HCO$^+$ gas gap is shallower than that of the continuum gap D2, which is 16, by a factor of two to four \citep{Pinte16}. 
This trend, the shallower gas gap than the dust gap, is consistent with the theoretical expectation for gaps opened by planets \citep{Fouchet07, Picogna15, Zhang18}.
Nevertheless, the angular resolution of our HCO$^+$ data is a factor of three coarser, and the error bars of our measurements are relatively large, compared to the continuum data in \citet{ALMA15}. 
Thus, it is not straightforward to compare the relative position and depth between the HCO$^+$ gas gap and the dust gap.

If the HCO$^+$ gas gap at a radius of 30 au in the HL Tau disk is opened by a gas giant planet, 
the planet mass is estimated to be 0.5--0.8 $M_{\rm J}$ from the gap depth of 4--8 with the stellar mass of HL Tau of 2.1 $M_\sun$, a gas temperature of 60 K, and a viscosity $\alpha$ of 10$^{-3}$ using the formula in \citet{Kanagawa15}.
Here, the gas temperature is adopted to be the estimated $T_{\rm ex}$ of the HCO$^+$ lines, assuming that HCO$^+$ is thermalized.
We note that the formula of planet mass and gap depth in \citet{Kanagawa15} is derived for vertically isothermal disks. 
In the HL Tau disk, there is a vertical temperature gradient. 
At the radius of 30 au, the mid-plane temperature is likely 50 K, traced by the continuum emission, and the surface temperature is likely 100 K, which is the CO (1--0) brightness temperature.
In this formula derived by \citet{Kanagawa15}, the estimated planet mass is proportional to $T^{1.25}$, where $T$ is the gas temperature.

The estimated mass of the putative planet to open the HCO$^+$ gas gap at a radius of 30 au, 0.5--0.8 $M_{\rm J}$, is comparable to those estimates\footnote{We note that the stellar mass adopted in these simulations is different from our measurement of 2.1 $M_\sun$ from the Keplerian rotation observed in the HCO$^+$ (3--2) emission with ALMA. The stellar mass of 1 $M_\sun$ was adopted in \citet{Dong15}, 1.3 $M_\sun$ in \citet{Dipierro15} and 0.55 $M_\sun$ in \citet{Picogna15} and \citet{Jin16}. As shown in \citet{Kanagawa15}, derived planet mass is proportional to stellar mass. Thus, we linearly scaled the mass of planets in these simulations by the ratio of their adopted and our measured stellar masses.} of 0.4 $M_{\rm J}$ in the dusty smoothed particle hydrodynamic simulations by \citet{Dipierro15} and of 0.7 $M_{\rm J}$ in the hydrodynamic disk gas+dust simulations by \citet{Jin16}, which reproduce the features of rings and gaps observed in the continuum emission with ALMA. 
In \citet{Dong15} and \citet{Picogna15}, their simulations show that the gap at a radius of 30 au observed in the continuum with ALMA can be opened by a planet with a mass of 0.3--0.4 $M_{\rm J}$.
Nevertheless, these estimates are all well below the upper limit of the mass of the putative planet of 15 $M_{\rm J}$ set by the non-detection with the direct-imaging observations \citep{Testi15}.

At radii close to 70 au, two gaps, D5 at a radius of 64 au and D6 at a radius of 74 au, are detected in the continuum emission in the HL Tau disk \citep{ALMA15}, 
and their depths are estimated to be 8 and 12, respectively \citep{Pinte16}.
If the D5 and D6 gaps are considered as a single gap with remnant dust inside, 
the gap depth is estimated to be 16 \citep{Pinte16}.
Numerical simulations suggest that a planet with a mass of 0.9--1.4 $M_{\rm J}$ \citep{Picogna15,Dipierro15,Jin16} or 0.4 $M_{\rm J}$ \citet{Dong15} can open a gap at a radius close to 70 au in a disk similar to the HL Tau disk.
Our estimated radial profile of the HCO$^+$ column density does not show a clear gap at a radius of 60--70 au. 
There is a hint of a decrease in the HCO$^+$ column density at the radius of the continuum D5 gap of 64 au.
The depth of this decrease in the column density is less than three. 
With a depth of three, our estimated $T_{\rm ex}$ of the HCO$^+$ lines at that radius, and the formula by \citet{Kanagawa15}, 
the mass of the putative planet to open the shallow HCO$+$ gas gap (if present) at a radius of 64 au is derived to be 0.2 $M_{\rm J}$, 
on the assumption that the HCO$^+$ abundance is constant in the HL Tau disk.
This value is lower than those adopted in the numerical simulations \citep{Dong15,Picogna15,Dipierro15,Jin16}.
With a planet having a mass of 0.4 $M_{\rm J}$ or 1 $M_{\rm J}$ as those adopted in the numerical simulations, 
we expect a deeper gap with a depth of five or even larger than ten in gas.
Such a deep gas gap at a radius close to 70 au is not seen in our HCO$^+$ data.

\begin{figure}
\centering
\includegraphics[width=8cm]{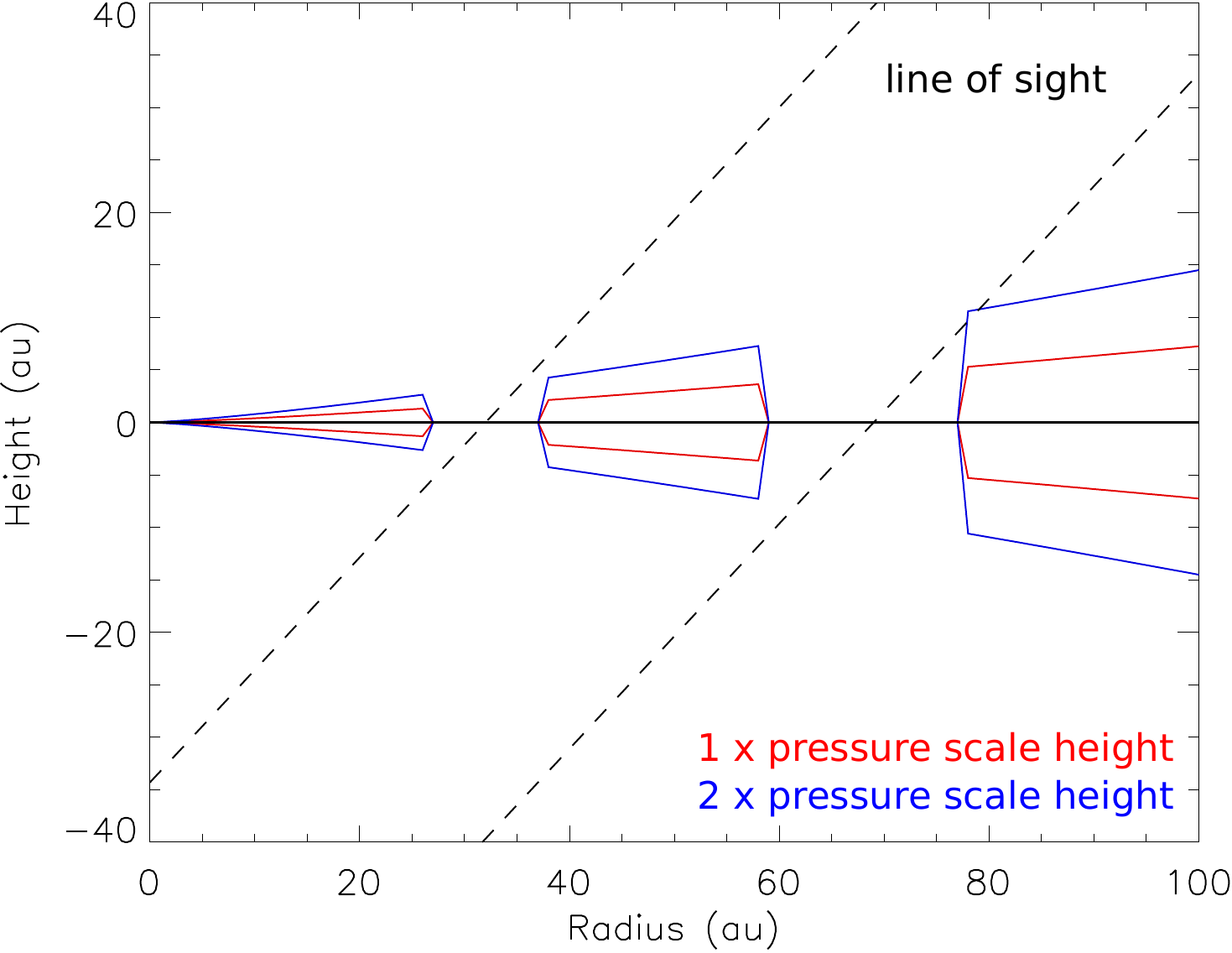}
\caption{Schematic figure of the flared disk around HL Tau. A black horizontal line shows the disk mid-plane. Red and blue lines delineate one and two times the pressure scale height of the HL Tau disk, respectively. Two gaps at radii of 30 and 70 au are shown in the figure. Dashed lines present the line of sight with an inclination angle of 47$\arcdeg$. The pressure scale height is computed with the estimated excitation temperature of the HCO$^+$ lines, assuming that HCO$^+$ is thermalized. }
\label{hr}
\end{figure}

We note that the continuum emission most likely traces the mid-plane of the HL Tau disk \citep{Kwon11}, 
while the HCO$^+$ lines trace an upper layer \citep{Oberg11}.
The contrast in the column density at radii inside and outside a gap could be suppressed due to the projection effect 
because the gap could be obscured by the disk surface at adjacent radii.
Figure \ref{hr} presents a schematic plot of the pressure scale height as a function of radius in the HL Tau disk.
The pressure scale height is computed as $h(r) = r \times c_{\rm s}(r)/V_{\rm kep}(r)$, where $c_{\rm s}(r)$ is the sound speed and $V_{\rm kep}$ is the Keplerian velocity at a radius of $r$.
The sound speed is computed as  $\sqrt{kT_{\rm g}/\mu m_{\rm H}}$, where $k$ is the Boltzmann constant, $\mu$ is the mean molecular weight of 2.3, $m_{\rm H}$ is the atomic mass, and the gas temperature $T_{\rm g}$ is adopted to be $T_{\rm ex}$ of the HCO$^+$ lines on the assumption that the HCO$^+$ lines are thermalized.
If the HCO$^+$ lines trace the upper layer at one pressure scale height, 
the outer gap (if present) at radii of 64--74 au, the same as the continuum D5 and D6 gaps, is not obscured by the disk surface at the adjacent radii.
However, if the HCO$^+$ emission originates from an upper layer with height more than twice the pressure scale height, corresponding to a disk aspect ratio ($h/r$) of 0.13, 
the outer gap could be obscured because of the projection effect.
Future observations in optically-thinner molecular lines to probe a layer closer to the mid-plane are required to examine the presence of the outer gap.

\subsubsection{Gap formed by chemical effects?}
An HCO$^+$ gas gap could form in a protoplanetary disk due to the CO depletion, even if there is no deficit of H$_2$ gas at a certain radius.
HCO$^+$ forms via chemical reactions from CO. 
When CO is absorbed onto dust grains in dense regions with a temperature lower than 20--30 K \citep{Furuya14}, 
the HCO$^+$ abundance also decreases. 
Consequently, an HCO$^+$ gas gap forms at an inner radius, where CO is depleted, in a protoplanetary disk \citep[e.g.,][]{Walsh13, Aikawa15}. 
In the HL Tau disk, the dust temperature at a radius of 30 au is above 30 K, where the HCO$^+$ gas gap is present \citep{Okuzumi16, Pinte16}.
Thus, the presence of the HCO$^+$ gas gap at a radius of 30 au in the HL Tau disk is not caused by the CO depletion.

In addition to the CO depletion, an HCO$^+$ gas gap could form when H$_2$ ionization rate at outer radii in a protoplanetary disk is enhanced. 
Chemical models of protoplanetary disks show that when the cosmic-ray ionization rate in the mid-plane at outer radii is on the order of 10$^{-17}$ s$^{-1}$, 
the HCO$^+$ abundance and thus the HCO$^+$ column density at outer radii increase \citep{Cleeves14}. 
On the other hand, the HCO$^+$ column density decreases at inner radii, where the cosmic-ray ionization is attenuated or where CO is depleted. 
The increase and decrease in the HCO$^+$ column density at outer and inner radii result in an HCO$^+$ gas gap with a width of several tens of au \citep{Cleeves14}. 
The width of the observed HCO$^+$ gas gap in the HL Tau disk is narrower than 20 au. 
Such a narrow HCO$^+$ gas gap is unlikely formed by an enhancement of the H$_2$ ionization rate at outer radii.

The other possibility to reduce the HCO$^+$ abundance is excess of UV radiation \citep[e.g.,][]{Jonkheid07}. 
Because the observed HCO$^+$ gas gap at a radius of 30 au coincides with the dust gap, 
there is a deficit of dust grains in the HCO$^+$ gas gap \citep{Pinte16, Okuzumi16}. 
Thus, the gas-to-dust mass ratio is higher in the gap, if there is no actual deficit of H$_2$ gas.
As demonstrated in chemical models, 
the HCO$^+$ abundance in a protoplanetary disk with a higher gas-to-dust mass ratio of 1000 could be a factor of ten lower than in a disk with a gas-to-dust mass ratio of 100 \citep{Jonkheid07}. 
Therefore, the HCO$^+$ column density possibly decreases in these dust gaps, even if there is no deficit of H$_2$ gas in the dust gaps.
Future observations to measure column density of neutral molecules in the HL Tau disk, whose abundances are less sensitive to the gas-to-dust mass ratio and UV radiation, are needed to examine this possibility.  

\subsection{Origins of the infalling spiral}
An infalling gas stream toward HL Tau is detected in the HCO$^+$ (3--2) emission with our ALMA observations, 
and its rotational motion gradually becomes dominant over its infalling motion.
We note that the HCO$^+$ (3--2) line is optically thick (Fig.~\ref{hcop_tau}), and the observed line-of-sight velocity of the spiral structure is inconsistent with the Keplerian velocity at radii larger than 0\farcs3 (Fig.~\ref{spiral_pv}).
In addition, no counterpart of this spiral is detected in the continuum emission, which traces the disk mid-plane.
These results could suggest that the infalling spiral is not in the mid-plane but is located above the disk surface. 
The velocity of the infalling spiral matches with the Keplerian velocity at a radius of 0\farcs3, 
and there is a bright intensity peak at a radius close to 0\farcs3 (Fig.~\ref{hcop_mom}). 
Thus, the infalling spiral could merge with the HL Tau disk at a radius of 0\farcs3.
Future observations of shock tracers, such as SO, SO$_2$, and CH$_3$OH, are needed to examine this scenario and the presence of accretion shock \citep{Ohashi14, Sakai14, Yen14, Villarmois19}.

Spiral structures connected with central disks have been observed in millimeter continuum emission and infrared reflected light around several young stellar objects \citep{Liu16, Liu18}.  
Infalling motion along spiral structures have also been detected around other protoplanetary disks \citep{Tang12, Yen14}.
An infalling spiral could form when a gas clump free falls towards the center with a conserved angular momentum \citep[e.g.,][]{Ulrich76}.
Free-falling material with a conserved angular momentum follows a parabolic trajectory, 
which appears as a spiral structure after the projection on the plane of the sky. 
This free-fall model has successfully explained the morphology and velocity of the infalling spirals observed in the protostellar source, L1489 IRS \citep{Yen14}.
In HL Tau, we found that the morphology of the outer part of the infalling spiral at a radius larger than 0\farcs5 indeed can be described by a parabolic trajectory projected on the plane of the sky, but that of the inner part cannot because its curvature is too large for a projected parabolic trajectory.
In addition, the observed velocity along the infalling spiral in HL Tau cannot be explained with a free-fall motion with a conserved angular momentum, as discussed in Section \ref{spiral}.
Thus, the infalling spiral observed in HL Tau is unlikely formed by a gas clump free falling toward the disk.

If the density structures in a collapsing dense core are asymmetric, it is natural to form infalling spirals.
As demonstrated in numerical simulations of collapse of dense cores in magnetized turbulent giant molecular clouds,  
density structures on a scale of thousands of au are typically filamentary, 
and central disks could be connected with spiral-like infalling gas streams \citep{Kuffmeier17}.
The protostellar envelope around HL Tau indeed exhibits asymmetric structures on a 1000 au scale \citep{Yen17, Wu18},
and it is embedded in a filamentary structure on a 0.1 pc scale \citep{Welch00, Yen19}.
In addition, the envelope is likely impacted by a large-scale expanding shell driven by the wide-angle outflow or wind from the nearby young star XZ~Tau \citep{Yen19}. 
Therefore, the asymmetric structures in the envelope could lead to the formation of the infalling spiral connected with the disk in HL Tau.

Furthermore, infalling spirals could form even if the initial density distributions of a collapsing dense core is symmetric. 
In magnetohydrodynamic (MHD) simulations of turbulent dense cores, 
the pseudo disk is warped by the turbulence, resulting in spiral structures around a central rotationally supported disk \citep{Li14}.
Different from the HL Tau disk, which is linked with one prominent spiral structure, 
in these simulations, multiple spirals form and connect to the disk \citep[Fig.~2 in][]{Li14}.
Infalling spirals also form in MHD simulations without incorporating turbulence.
In MHD simulations with enhanced ambipolar diffusion by removing small dust grains with sizes smaller than 0.1 $\mu$m, 
a rotationally supported disk connected with infalling spirals could form because of less efficient magnetic braking \citep{Zhao16, Zhao18}.
In these simulations, the infalling material with a high specific angular momentum forms a ring-like structure at its centrifugal radius, which surrounds the central disk. 
Then this ring-like structure breaks into one or two prominent spirals, which are connected to the central disk. 
These spirals are located in the regions where gas flows converge.
In these spirals, the rotational velocity is higher than the Keplerian velocity, and the infalling velocity becomes lower than the rotational velocity, at radii close to the disk \citep[Fig.~14 and 15 in][]{Zhao18}.
Therefore, the properties of the observed infalling spiral in HL Tau are qualitatively similar to those in the MHD simulations with less efficient magnetic braking due to enhanced ambipolar diffusion.
Adding data from ALMA observations with shorter baselines is needed to recover the motion and structure of the flattened envelope around the HL Tau disk at a high angular resolution and to further study the origin of the infalling spiral.

\section{Summary}
We present our observational results of the 1.1 mm continuum and the HCO$^+$ (3--2) emission in HL Tau at angular resolutions of 0\farcs1 obtained with ALMA and our data analysis of the 2.9 mm and 1.1 mm continuum and the HCO$^+$ (3--2) and (1--0) lines of the HL Tau disk. 
Our main results are summarized below.

\begin{enumerate}
\item{The HCO$^+$ (3--2) emission traces the HL Tau disk. 
The Keplerian rotation of the disk is clearly resolved. The stellar mass of HL Tau is measured to be 2.1$\pm$0.2 $M_\sun$ with an inclination angle of the disk mid-plane of 47$\arcdeg$. 
In addition, there is a bright peak in the HCO$^+$ (3--2) emission located 0\farcs1 north of the center. 
A one-arm spiral structure stretches out from that intensity peak toward the northwest, and then bends toward the southwest. 
The spiral extends to 2$\arcsec$ away from the disk in the southwest. 
Deep absorption is also observed in the southwestern part of the disk in the HCO$^+$ line, 
which is possibly caused by the infalling envelope in front of the disk.}

\item{We measured the radial profiles of the HCO$^+$ column density and excitation temperature from the integrated intensity of the HCO$^+$ (3--2) emission and the intensity ratio of the HCO$^+$ (3--2) to (1--0) emission. 
The estimated excitation temperature of the HCO$^+$ lines ranges from 25 K to 65 K at radii from 15 au to 130 au, 
which is 5 K to 10 K higher than the dust temperature and a factor of two to three lower than the CO (1--0) brightness temperature.
An HCO$^+$ gas gap at a radius of 30 au, where the column density drops by a factor of 4--8, is seen in the column density profile, 
and it is coincident with a dust gap.
There is no other clear HCO$^+$ gap seen at outer radii.
The absence of HCO$^+$ gaps at outer radii could be due to the projection effect, if the HCO$^+$ emission originates from an upper layer with height more than twice the pressure scale height, corresponding to a disk aspect ratio larger than 0.13.}

\item{The HCO$^+$ gas gap at a radius of 30 au observed in the HL Tau disk could be opened by a gas giant planet with a mass of 0.5--0.8 $M_{\rm J}$, 
which is comparable to the planet mass adopted in numerical simulations to form a dust gap similar to the observations at that radius.
This HCO$^+$ gas gap is unlikely caused by CO depletion or a change in the ionization rate 
because of the high dust temperature above 30 K at a radius of 30 au and the narrow width of the HCO$^+$ gas gap.
Nevertheless, the HCO$^+$ abundance could be preferentially lower in the dust gap because more UV radiation penetrates through the dust gap. 
Future observations to measure column density of neutral molecules in the HL Tau disk, whose abundances are less sensitive to the gas-to-dust mass ratio and UV radiation, are needed to confirm the gas gap and to measure the gap depth more robustly.}

\item{The spiral structure is observed at redshifted velocities. 
At radii larger than 0\farcs3, its line-of-sight velocity is more redshifted than the expectation from the Keplerian rotation, 
and the velocity of the spiral matches with the Keplerian velocity at radii smaller than 0\farcs3.
Because of the excess of the redshifted velocity at outer radii, where the near side of the envelope is, 
the spiral is likely infalling toward the disk. 
On the other hand, at radii smaller than 0\farcs6, the spiral structure is located in the northwest of HL Tau at its far side.
The infalling and rotational motions there induce blueshifted and redshifted velocities, respectively.  
The redshifted velocity of the spiral structure on the far side suggests that the rotational motion dominates over the infalling motion at these smaller radii.
Nevertheless, the radial profiles of the infalling and rotational velocities of the spiral structure cannot be fully determined because the 3D orientation of the spiral is not known.
}

\item{The infalling spiral toward the HL Tau disk is likely located above the disk surface.  
The formation of this infalling spiral could be related to the asymmetric density structures in the infalling protostellar envelope on a 1000 au scale around HL Tau. 
As shown in numerical simulations, asymmetric collapsing dense cores tend to form spiral-like infalling gas streams around central disks.
In addition, such an infalling spiral with its rotational velocity exceeding the Keplerian velocity and dominant over the infalling velocity at radii close to a central disk has been seen in MHD simulations of collapsing dense cores with less efficient magnetic braking due to enhanced ambipolar diffusion.}

\end{enumerate}

\begin{acknowledgements} 
We thank Yuri Aikawa and Bo Zhao for useful discussions and help.
This paper makes use of the following ALMA data: ADS/JAO.ALMA\#2016.1.00366.S and ADS/JAO.ALMA\#2011.0.00015.SV. ALMA is a partnership of ESO (representing its member states), NSF (USA) and NINS (Japan), together with NRC (Canada), MOST and ASIAA (Taiwan), and KASI (Republic of Korea), in cooperation with the Republic of Chile. The Joint ALMA Observatory is operated by ESO, AUI/NRAO and NAOJ. 
We thank all the ALMA staff supporting this work. 
H.W.Y. acknowledges support from MOST 108-2112-M-001-003.
P.M.K. acknowledges support from MOST 107-2119-M-001-023 and from an Academia Sinica Career Development Award.
S.T. acknowledges a grant from JSPS KAKENHI Grant Number JP18K03703 in support of this work. 
This work was supported by NAOJ ALMA Scientific Research Grant Numbers 2017-04A. 
\end{acknowledgements} 

\begin{appendix}

\begin{figure}
\centering
\includegraphics[width=9.2cm]{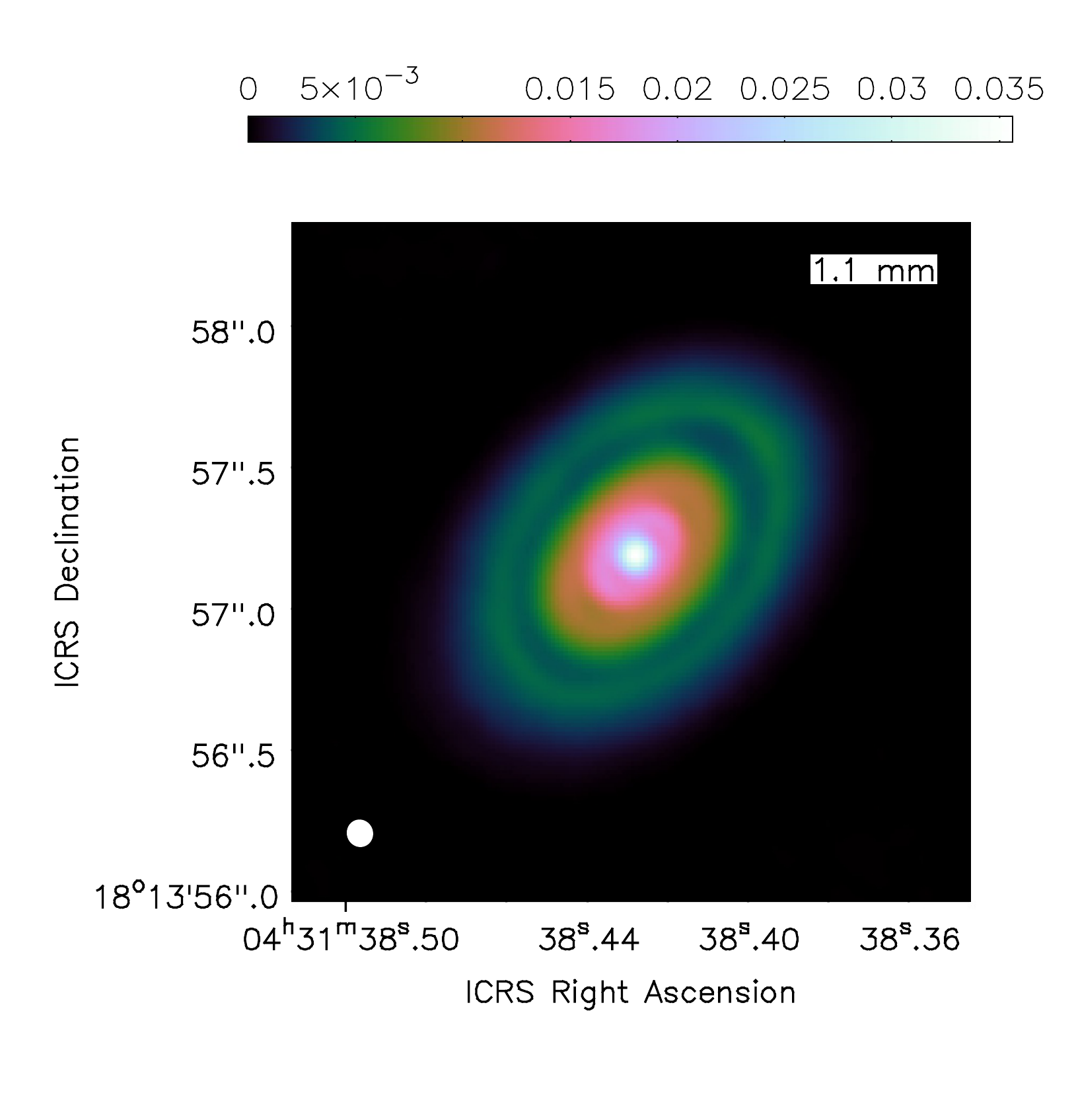}
\caption{1.1 mm continuum image of HL Tau obtained with our ALMA observations. Color scale is in units of Jy beam$^{-1}$. A white filled ellipse presents the size of the synthesized beam.}
\label{cont_im}
\end{figure}

\section{1.1 mm continuum}\label{1.1mm_cont}
Figure \ref{cont_im} presents the 1.1 mm continuum image of HL Tau obtained with our ALMA observations. 
The angular resolution of this 1.1 mm continuum image is a factor of three coarser than those of the continuum images at 1.3 mm and 0.87 mm presented in \citet{ALMA15}. 
The observed 1.1 mm continuum emission traces the protoplanetary disk around HL Tau. 
The overall morphology is consistent with that in the 2.9 mm, 1.3 mm and 0.87 mm continuum images in \citet{ALMA15}, 
and the gaps at radii of $\sim$30~au and $\sim$60--70~au are clearly seen in this 1.1 mm continuum image.
The peak position of the 1.1 mm continuum emission is measured to be 04$^{\rm h}$31$^{\rm m}$38\fs428 +18$\arcdeg$13$\arcmin$57\farcs19 with our observations in 2017. 
Because of the proper motion of HL Tau \citep{Zacharias04}, 
the continuum peak position measured in 2017 is different from that measured by \citet{ALMA15} in 2014, which is 04$^{\rm h}$31$^{\rm m}$38\fs425 +18$\arcdeg$13$\arcmin$57\farcs24.

We measured the total flux of the 1.1 mm continuum emission of the disk around HL Tau using the method of the curve of growth \citep[e.g.,][]{Ansdell16}.
We set an aperture with an aspect ratio of 0.68 and a PA of 138$\arcdeg$ centered at the position of HL Tau, which corresponds to the geometry of the circumstellar disk around HL Tau projected on the plane of the sky. 
We gradually increased the size of the aperture and measured the flux enclosed in the aperture until the measured flux reached its maximum. 
The 1.1 mm flux of the disk is measured to be 894.4$\pm$0.4 mJy.
The uncertainty is estimated as $\sqrt{A_{\rm disk}/A_{\rm beam}}\times \sigma$, where $A_{\rm disk}$ and $A_{\rm beam}$ are the area of the disk and the synthesized beam, $\sqrt{A_{\rm disk}/A_{\rm beam}}$ corresponds to the number of independent measurements, and $\sigma$ is the noise level of 30 $\mu$Jy. 
The radius enclosing 95\% of the total flux is estimated to be 0\farcs82$\pm$0\farcs02 (115$\pm$3 au).

The spectral index of the continuum emission between 1.3 mm (233 GHz) and 0.87 mm (343.5 GHz) averaged over the entire disk is 2.6 \citep{ALMA15}.
By interpolating these measurements, 
the continuum flux at 1.1 mm is 997 mJy, 
which is consistent with our observational measurement at 1.1 mm within 10\%. 
Thus, the flux measured in our observations is consistent with the previous ALMA observations within the uncertainty of an absolution flux calibration of 5--10\%.
\end{appendix}

\software{CASA \citep[v4.7.0;][]{McMullin07}}
\software{RADEX \citep{radex07}}
\software{MPFIT \citep{Markwardt09}}

\end{document}